\documentclass[12pt, draftclsnofoot, onecolumn]{IEEEtran}
\usepackage{geometry}
\ifCLASSINFOpdf

\else

\fi
\usepackage[cmex10]{amsmath}
\usepackage{amssymb}
\usepackage{cite}
\usepackage{graphicx}
\usepackage{array,color}
\usepackage{amsmath}
\usepackage{stfloats}
\usepackage{graphicx}
\usepackage{subfigure}
\usepackage{tabularx}
\usepackage{epsfig,epsf,color,balance,cite}
\usepackage{algorithmic}
\usepackage{algorithm}
\usepackage{bm}
\usepackage{textcomp}

\begin{document}

\title{ Intelligent Reflecting Surface Aided MIMO Broadcasting for Simultaneous Wireless Information and Power Transfer}
\author{ Cunhua Pan, Hong Ren, Kezhi Wang,  Maged Elkashlan,  Arumugam Nallanathan, \IEEEmembership{Fellow, IEEE},  Jiangzhou Wang, \emph{Fellow}, \emph{IEEE} and Lajos Hanzo, \IEEEmembership{Fellow, IEEE}
%\thanks{This work was supported by ...}
\thanks{C. Pan, H. Ren, M. Elkashlan and A. Nallanathan are with the School of Electronic Engineering and Computer Science at  Queen Mary University of London, London E1 4NS, U.K. (e-mail:\{c.pan, h.ren, maged.elkashlan, a.nallanathan\}@qmul.ac.uk). K. Wang is with Department of Computer and Information Sciences, Northumbria University, UK. (e-mail: kezhi.wang@northumbria.ac.uk). J. Wang is with the School of Engineering and Digital Arts, University of Kent, Canterbury, Kent, CT2 7NZ, U.K. (e-mail: J.Z.Wang@kent.ac.uk). L. Hanzo is with the School of Electronics and Computer Science, University of Southampton, Southampton, SO17 1BJ, U.K. (e-mail: lh@ecs.soton.ac.uk). }
}

\maketitle
\vspace{-1.9cm}
\begin{abstract}
An intelligent reflecting surface (IRS) is invoked for enhancing  the energy harvesting performance of a simultaneous wireless information and power transfer (SWIPT) aided system. Specifically, an IRS-assisted SWIPT  system is considered, where  a multi-antenna aided base station (BS) communicates with several multi-antenna assisted information receivers (IRs), while guaranteeing the energy harvesting requirement of the energy receivers (ERs). To maximize the weighted sum rate (WSR) of IRs, the transmit precoding (TPC) matrices of the BS and passive phase shift matrix of the  IRS should be jointly optimized. To tackle this challenging optimization problem, we first adopt the classic  block coordinate descent (BCD) algorithm for decoupling the original optimization problem into several subproblems and alternatively optimize the TPC matrices and the phase shift matrix. For each subproblem, we provide a low-complexity iterative algorithm, which is guaranteed to converge to the Karush-Kuhn-Tucker (KKT) point of each subproblem. The BCD algorithm is rigorously proved to converge to the KKT point of the original problem.  We also conceive a feasibility checking method to study its feasibility. Our extensive simulation results confirm that employing IRSs in SWIPT beneficially enhances the system performance and the proposed BCD algorithm converges rapidly, which is appealing for practical applications.
\end{abstract}
\begin{IEEEkeywords}
Intelligent Reflecting Surface (IRS), Large Intelligent Surface (LIS), SWIPT, Energy Harvesting, MIMO.
\end{IEEEkeywords}

\IEEEpeerreviewmaketitle
%\newpage
\section{Introduction}
Recently, intelligent reflecting surface (IRS)-assisted wireless communication has received considerable research attention, since   it is capable of supporting  cost-effective and energy-efficient  high data rate communication for  next-generation communication systems \cite{di2019smart,qingqing2019towards,zhang2019multiple}. In simple tangible terms, an IRS is  composed of a vast number of low-cost and passive reflective components, each of which is capable of imposing a phase change on the  signals incident upon them. Thanks to the recent advances in  meta-materials \cite{cui2014coding}, it has become feasible to reconfigure the phase shifts in real time. As a result, the phase shifts of all reflective components can be  collaboratively adjusted for ensuring that the signals  reflected from the IRS can be added constructively or destructively at the receiver in order to beneficially steer the  signal component arriving from the base station (BS) for enhancing the desired signal power or alternatively for suppressing the undesired signals, such as interference. In contrast to conventional physical layer techniques that are designed for accommodating the hostile time-varying wireless channels \cite{zhang2019cell,zhang2018mixed}, IRSs constitute a new paradigm capable of `reprogramming' the wireless propagation environment into a more favorable transmission medium. Since the reflective components are passive, they impose a much lower power consumption than conventional relay-aided communication systems relying on active transmission devices. Additionally,   no thermal noise is imposed by the IRS, since it directly reflects the incident signals without decoding or amplifying them, which is in contrast to conventional relays. Furthermore, the reflective phase arrays can be fabricated in small size and low weight, which enables them to be easily coated in the buildings' facade, ceilings, walls, etc.  Furthermore, as IRS is a complementary device, it can be readily integrated into current wireless networks without modifying the physical layer standardization, making it transparent to the users. To fully exploit the benefits of IRS,   the active beamforming at the BS and the passive beamforming at the IRS should be jointly designed. However, the optimization variables are coupled and the joint design leads to a complex optimization problem that is difficult to solve.

Some innovative efforts have been devoted to the transceiver design when integrating IRS into various wireless communication systems, including the single-user scenarios of \cite{qingqingwuglobe,yu2019miso,yang2019intelligent,han2019large,abeywickrama2019intelligent}, the downlink multiple-user scenarios of \cite{wu2018intelligent,huang2019reconfigurable,guo2019weighted,nadeem2019large}, the  physical layer security design of \cite{yu2019enabling,cui2019secure,hongshen,chen2019intelligent,xu2019resource,guan2019intelligent}, the mobile edge computing (MEC) networks of \cite{tong2019}, multigroup multicast networks of \cite{gui2019} and the multicell multiuser multiple-input multiple-output (MIMO) case in \cite{pan2019intelligent}. Concretely, Wu \emph{et al.} proposed   joint active and passive beamforming for a single-user scenario in \cite{qingqingwuglobe}, where semidefinite relaxation (SDR) was proposed for optimizing the phase shift matrix. However, its complexity is high since the number of optimization variables increases quadratically with the number of phase shifts. Additionally, the Gaussian random approximation employed leads to certain performance loss. To resolve this issue, Yu \emph{et al.} \cite{yu2019miso} proposed a pair of efficient algorithms termed as fixed point iteration and manifold optimization techniques, which can guarantee  locally optimal solutions. As a further advance, the authors of \cite{yang2019intelligent}  considered realistic  frequency-selective channels. The phase shift design was studied in \cite{han2019large} when only statistical channel state information (CSI) is available. A sophisticated  phase shift model was derived in \cite{abeywickrama2019intelligent}, by taking into account a realistic amplitude-phase relationship. For the multiuser case, the authors in \cite{wu2018intelligent}  considered the total transmit power minimization problem, while guaranteeing the users' signal-to-interference-plus-noise ratio (SINR) constraints. The associated energy efficiency maximization problem was studied in \cite{huang2019reconfigurable} and zero-forcing beamforming was adopted by the BS for simplifying the optimization problem.  By contrast, a weighted sum rate (WSR) maximization problem was considered in \cite{guo2019weighted} and  the fairness issues were studied in \cite{nadeem2019large}.  The authors of \cite{yu2019enabling,cui2019secure,hongshen} studied the security issues  of a single-user case, while  the authors of \cite{chen2019intelligent,xu2019resource,guan2019intelligent} considered multiple-user scenarios. In \cite{tong2019}, the IRS was shown to be beneficial in reducing the latency of MEC networks. In addition, the IRS can help enhance the WSR performance for the multigroup multicast network in \cite{gui2019}.  Most recently, we considered an IRS-assisted multicell MIMO communications scenario \cite{pan2019intelligent}, where we demonstrated that deploying an IRS at the cell edge is also capable of mitigating  the   adjacent-cell interference. Channel state information (CSI) is challenging to obtain in IRS-assisted communication system due to its passive feature. There are some initial efforts to handle this issue such as channel estimation and/or robust transmission design  \cite{huang2019indoor,he2019cascaded,taha2019enabling,zhou2019robust}. Specifically, Huang \emph{et al.} \cite{huang2019indoor} proposed a deep learning method for efficient online configuration of the phase shifts, where the phase values can be immediately obtained by inputting the user location into the trained deep neural network. A two-stage channel estimation method based on a sparse matrix factorization and a matrix completion was proposed in \cite{he2019cascaded}. A pair of algorithms based on compressed sensing and deep learning were conceived by Taha \emph{et al.} \cite{taha2019enabling} for tackling the challenging channel estimation issues of IRS-assisted systems. Most recently, we first studied the robust beamforming design for IRS-assisted communication systems in \cite{zhou2019robust}, where the imperfect channel from an IRS to users was considered and the channel estimation error was assumed to be within a bounded elliptical region.

 \begin{figure}
\centering
\includegraphics[width=3.5in]{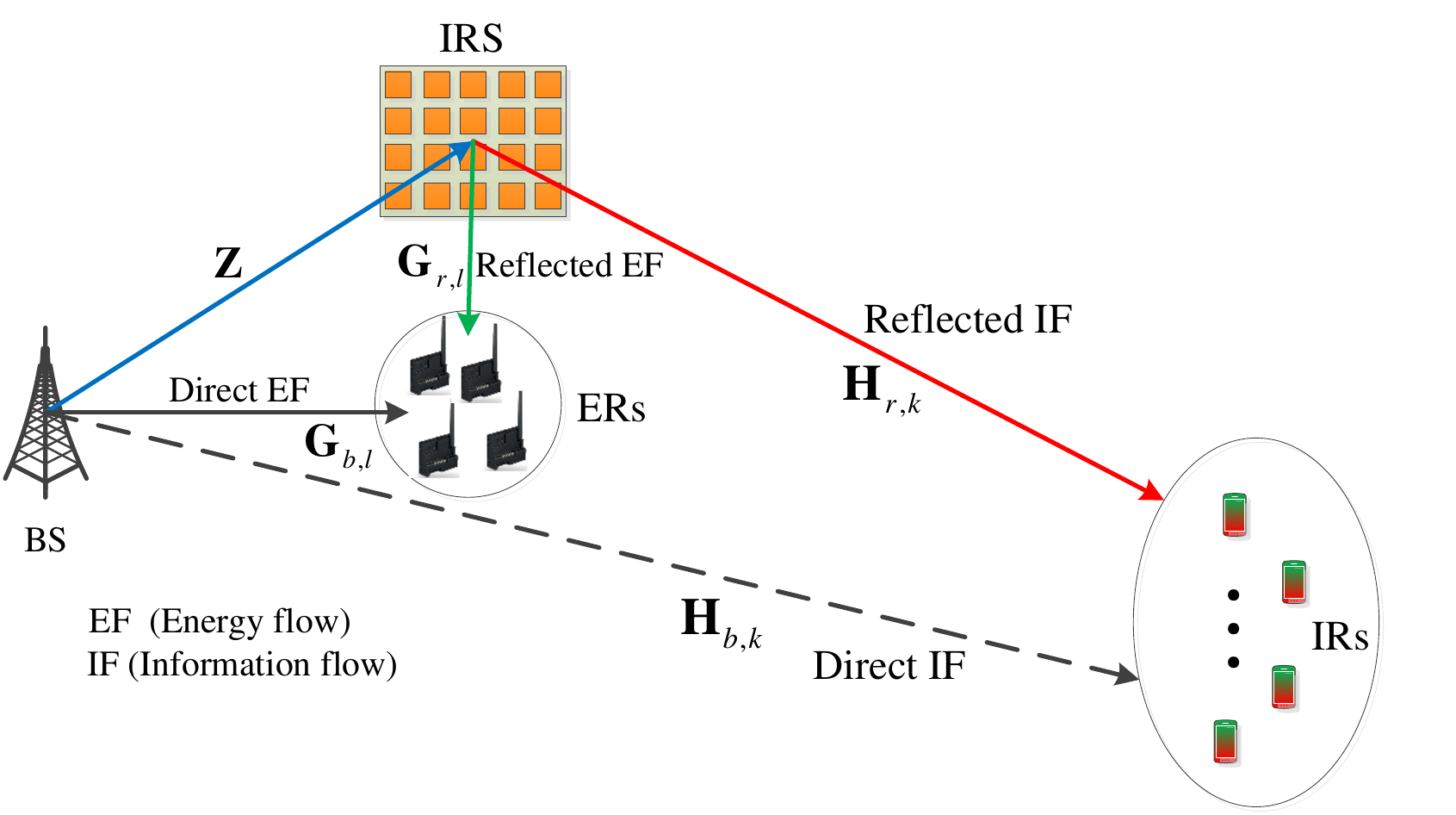}\vspace{-0.3cm}
\caption{An IRS-assisted SWIPT system. }\vspace{-0.9cm}
\label{fig1}
\end{figure}

On the other hand, information transmission enabled simultaneous wireless information and power transfer (SWIPT) is an appealing technique for future energy-hungry Internet-of-Things (IoTs) networks. Specifically,  a base station (BS) with constant power supply will transmit wireless signals to a set of devices. Some devices intend to decode the information from the received signal, which are termed as  information receivers (IRs), while the others will harvest the signal energy, which are called energy receiver (ER).
In  \cite{zhang2013mimo}, Zhang \emph{et al.} studied the trade-off between the information rate attained and the amount of harvested energy for a single-user MIMO system.  In practice, a typical ER such as a humidity sensor requires much higher  energy for its operation than that required by IRs. Due to severe channel attenuation, the power received  by the ERs is weak, which limits the maximum link-distance of ERs. To mitigate this issue, we propose to deploy an IRS in the vicinity of ERs to provide additional transmission links to support the ERs for enhancing their harvested power as shown in Fig.~\ref{fig1}, since there is a paucity of IRS-assisted SWIPT contributions in the literature \cite{wu2019weighted}.  Explicitly in \cite{wu2019weighted}, the weighted sum power maximization problem was studied by Wu and Zhang, who proved that no dedicated energy-carrying signals were   required for an IRS-aided SWIPT system. The SDR method was   adopted for solving the optimization problem, which exhibits a high computational complexity as well as imposing a performance degradation due to the associated rank-one extraction. However, this method is not applicable when each user is equipped with multiple antennas. Hence, in this paper we formulate a weighted sum rate (WSR) maximization problem for the IRS-assisted SWIPT MIMO system of Fig.~\ref{fig1}, in which an IRS is installed in the vicinity of ERs for compensating the associated power loss, while maximizing the WSR of distant IRs with the aid of passive beamforming.

Against this background, the main contributions of this paper are summarized as follows:
\begin{enumerate}
  \item  We formulate the WSR maximization problem by jointly optimizing the transmit precoding (TPC) matrices of the BS and those of the passive beamforming at the IRS for our IRS-assisted SWIPT MIMO system subject to a non-convex unit-modulus constraint imposed on the phase shifts, while simultaneously satisfying the energy harvesting requirement of the ERs. To the best of our knowledge, this is the first treatise  considering the WSR maximization problem of IRS-assisted SWIPT MIMO systems, which is much more challenging than the weighted sum power minimization problem of \cite{wu2019weighted} since the latter can be readily transformed into a convex optimization problem.  In contrast to the multicell system of \cite{pan2019intelligent}, an additional energy harvesting constraint is also imposed in our current study, which further complicates the analysis. Specifically, this constraint is non-convex and the optimization problem may become infeasible.   The WSR maximization problem is challenging to solve, since the optimization variables are highly coupled and the data rate expressions of the IRs are complex. To deal with this issue, we first reformulate the original problem into an equivalent form by exploiting the equivalence between the data rate and the weighted minimum  mean-square error (WMMSE). Then, an alternating optimization algorithm based on the popular block coordinate descent (BCD) algorithm is proposed for alternatively updating the active TPC matrices of the BS and the phase shift matrix of the IRS, which is rigorously proved to converge to the  Karush-Kuhn-Tucker (KKT) point of the original optimization problem.
  \item For a given phase shift matrix, we then proceed by developing an iterative algorithm based on the  successive convex approximation (SCA) method and on the Lagrangian dual decomposition method to derive a nearly closed-form solution for the TPC matrices. A low-complexity bisection search method is proposed for finding the optimal dual variables. The solutions generated by our iterative algorithm are guaranteed to converge to the KKT point of the TPC optimization problem.
  \item For the given TPC matrices,   we formulate the phase shift optimization problem as a non-convex quadratically constrained quadratic program (QCQP) subject to an additional energy harvesting constraint by invoking some further matrix  manipulations. Then, a novel iterative algorithm based on the majorization-minimization (MM) algorithm \cite{yansun} and on the price-based method \cite{cunhua2016} is developed for solving the QCQP. We strictly prove that the final solution generated by the iterative algorithm is guaranteed to converge to the KKT point of the phase shift optimization problem.
   \item  The associated feasibility issue is also studied by formulating an alternative optimization problem  and an iterative algorithm is proposed for solving this problem.
   \item Extensive simulation results are provided for verifying the performance advantages of employing IRS in SWIPT in order to enhance the energy harvesting performance. It is shown that the operating range of the ERs can be dramatically expanded by placing IRSs in the ERs' vicinity. Furthermore, the BCD algorithm converges rapidly,  and it is eminently suitable for practical applications. Our simulation results also show that as expected, the path loss exponent substantially  affects the system's performance  and thus the location of the IRS should be carefully chosen.
\end{enumerate}

The remainder of this paper is organized as follows. In Section \ref{system}, we introduce the IRS-assisted SWIPT system model and our problem formulation. The detailed algorithms used for solving the optimization problem are presented in Section \ref{algo}. The feasibility issues of the original problem are discussed in Section \ref{frjgtoj}, followed by our extensive simulations and discussions  in Section \ref{simlresult}. Finally, our conclusions are provided in Section \ref{conclu}.

\emph{Notations}: For matrix $\bf{A}$, $\bf{A}^*$ and $\bf{A}^\star$ represent the conjugate operator and converged solution, respectively.  ${\rm{Re}}\{a\}$ represents the real part of a complex value $a$.  ${{\mathbb{ C}}^{M }}$ denotes the set of $M \times 1$ complex vectors. ${{\mathbb{E}}}\{\cdot\} $ denotes the expectation operation.  For  two matrices $\bf A$ and $\bf B$, ${\bf{A}} \odot {\bf{B}}$ represents Hadamard product of $\bf A$ and $\bf B$. ${\left\| {\bf{A}} \right\|_F}$, ${\rm{tr}}\left( {\bf{A}} \right)$ and $\left| {\bf{A}} \right|$  denote the  Frobenius norm, trace operation and determinant  of ${\bf{A}}$, respectively. $\nabla {f_{\bf{x}}}\left( {\bf{x}} \right)$ denotes the gradient of the function $f$ with respect to (w.r.t.) the vector ${\bf{x}} $. ${\cal C}{\cal N}({\bf{0}},{\bf{I}})$ represents a random vector following the distribution of zero mean and unit variance matrix. $\arg\{\cdot\}$ means the extraction of phase information. ${\rm{diag}}(\cdot)$ denotes the diagonalization operation. ${\left( \cdot \right)^{*}}$, ${\left( \cdot \right)^{\rm{T}}}$ and ${\left(  \cdot \right)^{\rm{H}}}$ denote the conjugate, transpose and Hermitian operators, respectively.  ${\rm{arg}}(\cdot)$ means the phase extraction  operation.

\vspace{-0.3cm}\section{System Model and Problem Formulation}\label{system}
\subsection{System  Model}

Consider the IRS-aided multiuser MIMO downlink of a SWIPT system operating over the same frequency band both for data and energy  transmission, as shown in Fig.~\ref{fig1}. Let us assume that there are $K_I$ IRs and  $K_E$ ERs, respectively. It is also assumed that the BS is equipped with $N_B\ge 1$ antennas, while each IR and ER is equipped with $N_I\ge 1$ and $N_E\ge 1$ antennas, respectively. Let us denote the sets of IRs and ERs as ${\cal K}_I$ and ${\cal K}_E$, respectively. In general, low-power sensors require a certain amount of power (e.g., 0.1 mW) for their real-time operation. Due to the associated severe channel attenuation, the sensors should be deployed sufficiently close to the BS, which limits their practical implementation. To resolve this issue, we propose to employ an IRS, which has $M$ reflective elements in the ERs' vicinity for extending the operational range of sensors, as shown in Fig. 1. Firstly, the IRS increases the energy harvested by the ERs, and additionally it also assists in enhancing the   signal strength for distant IRs through careful phase shift optimization.

The number of data streams destined for each IR is assumed to be $d$, satisfying $1\le d \le \min\{N_B,N_I\}$. The  signal transmitted from the BS is given by
\begin{equation}\label{rui}
  {{\bf{x}}} = \sum\limits_{k = 1}^{K_I} {{\bf{F}}_{k}{\bf{s}}_{k}},
\end{equation}
where ${\bf{s}}_{k}\in \mathbb{C}^{d\times 1}$ is the $(d \times 1)$-element data symbol vector designated for the $k$th IR satisfying $\mathbb{E}\left[ {{{\bf{s}}_{k}}{{{\bf{s}}_{k}^H} }} \right] = {\bf{I}}_d$ and $\mathbb{E}\left[ {{{\bf{s}}_{i}}{ {{\bf{s}}_{j}^H}}} \right] = {\bf{0}}, {\rm{for}}\  i \ne j$, while ${\bf{F}}_{k} \in \mathbb{C}^{N_B\times d}$ is the linear TPC matrix used by the BS for the $k$th IR.  Assuming non-dispersive narrow-band transmission, the baseband equivalent channels spanning from the BS to the IRS, from the BS to the $k$th IR, from the BS to the $l$th ER, from the IRS to the $k$th IR, and finally from the IRS to the $l$th ER are modelled by the matrices ${\bf{Z}}\in \mathbb{C}^{M\times N_B}$, ${\bf{H}}_{b,k}\in \mathbb{C}^{N_I\times N_B}$, ${\bf{G}}_{b,l}\in \mathbb{C}^{N_E\times N_B}$, ${\bf{H}}_{r,k}\in \mathbb{C}^{N_I\times M}$, and ${\bf{G}}_{r,l}\in \mathbb{C}^{N_E\times M}$, respectively.
Let us denote  the diagonal reflection-coefficient matrix at the IRS by ${\bm{\Phi}}  = {\rm{diag}}\left\{ {{e^{j{\theta _1}}}, \cdots ,{e^{j{\theta _m}}}, \cdots ,{e^{j{\theta _M}}}} \right\}$ \footnote{$j$ is the imaginary unit.}, where ${\theta _m} \in [0,2\pi ]$ is the phase shift of the $m$-th reflective element.
 Due to absorption and diffraction, the signal power that has been reflected multiple times is ignored.
As a result, the signal received at the $k$th IR is given by
\vspace{-0.3cm}
\begin{equation}\label{uhuhhu}
{{\bf{y}}_{I,k}}=\left( {{{\bf{H}}_{b,k}} + {{\bf{H}}_{r,k}}{\bf{\Phi}} {\bf{Z}}} \right){\bf{x}} + {{\bf{n}}_{I,k}},\vspace{-0.3cm}
\end{equation}
where   ${\bf{n}}_{I,k}$ is the $k$th IR's noise vector satisfying ${\cal C}{\cal N}\left( {{\bf{0}},\sigma_I^2{{\bf{I}}_{N_I}}} \right)$. Similarly, the  signal received at the $l$th ER is given by
\vspace{-0.3cm}
\begin{equation}\label{uaxsdehu}
{{\bf{y}}_{E,l}}=\left( {{{\bf{G}}_{b,l}} + {{\bf{G}}_{r,l}}{\bf{\Phi}} {\bf{Z}}} \right){\bf{x}} + {{\bf{n}}_{E,l}},\vspace{-0.2cm}
\end{equation}
where   ${\bf{n}}_{E,l}$ is the $l$th ER's noise vector obeying the distribution of ${\cal C}{\cal N}\left( {{\bf{0}},\sigma_E^2{{\bf{I}}_{N_E}}} \right)$.

We assume that  all the CSIs are perfectly known at the BS, and the BS is responsible for calculating the phase shifts of the IRS, which are then fed back by them to the IRS controller through dedicated feedback channels. Given this  idealized and simplified assumption, the results obtained represent a performance upper bound of how much performance gain can be achieved by an IRS.
Let us define  the equivalent channel spanning from the BS to the $k$th IR by $ {\bf{\bar H}}_{k} \buildrel \Delta \over =  {{{\bf{H}}_{b,k}} + {{\bf{H}}_{r,k}}{\bf{\Phi}} {\bf{Z}}}$. Upon substituting $\bf{x}$ into (\ref{uhuhhu}),
 ${{\bf{y}}_{I,k}}$ can be rewritten as
 \vspace{-0.2cm}
 \begin{equation}\label{uhsddhhu}
{{\bf{y}}_{I,k}} = {{{\bf{\bar H}}}_k}{{\bf{F}}_k}{{\bf{s}}_k} + \sum\limits_{i = 1,i \ne k}^{{K_I}} {{{{\bf{\bar H}}}_k}{{\bf{F}}_i}{{\bf{s}}_i}}  + {{\bf{n}}_{I,k}}.\vspace{-0.2cm}
\end{equation}
Then, the achievable data rate (nat/s/Hz) of the $k$th IR is given by \cite{pancunhua2017}
\vspace{-0.3cm}
\begin{equation}\label{hufasdxsdeweu}
{R_{k}}\left( {{\bf{F}},{\bm{\Phi}}  } \right) = {\log}\left| {{\bf{I}} + {{{\bf{\bar H}}}_{k}}{{\bf{F}}_{k}}{\bf{F}}_{k}^{\rm{H}}{\bf{\bar H}}_{k}^{\rm{H}}{\bf{J}}_{k}^{ - 1}} \right|,\vspace{-0.3cm}
\end{equation}
where ${\bf{F}}$ denotes the collection of TPC matrices, while ${{\bf{J}}_{k}}$ is the interference-plus-noise covariance matrix given by ${{\bf{J}}_{k}} = \sum\nolimits_{m = 1,m \ne k}^{K_I} {{{{\bf{\bar H}}}_{k}}{{\bf{F}}_{m}}{\bf{F}}_{m}^{\rm{H}}{\bf{\bar H}}_{k}^{\rm{H}}}   + {\sigma_I ^2}{\bf{I}}$.

On the other hand, due to the broadcast nature of wireless channels, the ERs can extract  energy from the electromagnetic wave. In general, the harvested power is nonlinear over the received radio frequency (RF) power  due to the nonlinear RF-to-DC conversion, which depends on the input RF power level. This nonlinear EH model has been characterized in \cite{boshkovska2015practical}, which is a complex function of the RF power. Based on this nonlinear EH model, various transmission designs have been proposed in \cite{mishra2018energy} and \cite{xiong2017rate}. However, there is still lack of a general model that can accurately characterize this nonlinear relationship by capturing all practical factors. Hence, for simplicity, we adopt the simple linear EH model as widely used in the existing literature \cite{zhang2013mimo,fangwang2018,xu2014multiuser}. By ignoring the noise power at the ERs, the total harvested power is proportional to the total received power. Let us define  the equivalent channel spanning from the BS to the $l$th ER by $ {\bf{\bar G}}_{l} \buildrel \Delta \over =  {{{\bf{G}}_{b,l}} + {{\bf{G}}_{r,l}}{\bf{\Phi}} {\bf{Z}}}$. Then, the total  power harvested by the $l$th ER is
\vspace{-0.2cm}
\begin{equation}\label{xsfrf}
  {Q_i} = \eta {\rm{tr}}\left( {\sum\limits_{k = 1}^{K_I} {{{{\bf{\bar G}}}_l}{{\bf{F}}_k}{\bf{F}}_k^{\rm{H}}{\bf{\bar G}}_l^{\rm{H}}} } \right),\vspace{-0.2cm}
\end{equation}
where $0<\eta\le 1$ is the energy harvesting efficiency.   In this paper, we consider the constraint that the weighted sum of the power harvested by all ERs  should be higher than a predefined value, which is
\begin{equation}\label{htyju}
 Q = \sum\limits_{l = 1}^{{K_E}} {{\alpha _l}{Q_l}}  = {\rm{tr}}\left( {\sum\limits_{k = 1}^{K_I} {{\bf{F}}_k^{\rm{H}}{\bf{G}}{{\bf{F}}_k}} } \right)\ge \bar Q,
\end{equation}
where ${\bf{G}} = \sum\nolimits_{l = 1}^{{K_E}} {{\alpha _l}\eta {\bf{\bar G}}_l^{\rm{H}}{{{\bf{\bar G}}}_l}} $, ${\alpha _l}$ is the energy weighting factor of the $l$th ER, with a higher value of ${\alpha _l}$ representing a higher priority for the $l$th ER than for others. Finally,  $\bar Q$ is the minimum harvested power threshold.

\vspace{-0.5cm}
\subsection{Problem Formulation}
\vspace{-0.2cm}
Upon introducing the notations of  $\phi _m={e^{j{\theta _m}}}, \forall m$, we have ${\bm{\Phi}}  = {\rm{diag}}\left\{ {\phi _1, \cdots ,   \phi _M} \right\}$. Again, we aim for jointly optimizing the TPC matrices ${\bf{F}}$ and phase shift matrix  ${\bm{\Phi}}$ with the goal of maximizing the WSR of all IRs subject to the total power budget, to the unit modulus of the phase shifters  and to the  harvested power requirement. Then, this problem can be formulated as follows:
\vspace{-0.6cm}
\begin{subequations}\label{appstaoneorig}
\begin{align}
\mathop {\max }\limits_{{{\bf{F}},{\bm{\Phi}}  } } \quad
&   {\sum\limits_{k = 1}^{K_I} {{\omega_{k}}{R_{ k}}\left( {{\bf{F}},{\bm{\Phi}} } \right)} }
\\
\qquad\ \textrm{s.t.}\qquad
&\sum\limits_{k = 1}^{K_I} {\left\| {{{\bf{F}}_{k}}} \right\|_F^2}  \le {P_{T}}, \label{dehwifr}\\
& {\rm{tr}}\left( {\sum\limits_{k = 1}^{K_I} {{\bf{F}}_k^{\rm{H}}{\bf{G}}{{\bf{F}}_k}} } \right)\ge \bar Q, \label{hhudiuwe}\\
& \left| {{\phi _m}} \right| = 1, m = 1, \cdots ,M,\vspace{-0.2cm} \label{jiofjj}
\end{align}
\end{subequations}
where $\omega_{k}$ is the weighting factor controlling the scheduling priority for each IR and ${P_{T}}$ is the power limit at the BS, while (\ref{jiofjj}) is the unit-norm constraint imposed on the phase shifters.

As the IRS is passive and both the ERs and IRs are energy constrained, we assume that this optimization problem is solved at the BS which posses the knowledge of the CSI of all related links  and other related parameters such as $\bar Q$. After computing the phase shift values for the IRS, they are sent to the IRS controller through dedicated control channels.   Problem (\ref{appstaoneorig}) is difficult to solve, since the TPC matrices and the phase shifts are coupled. If we remove the energy harvesting (EH) constraint, the problem reduces to the WSR maximization problem recently studied in \cite{pan2019intelligent}. However, the additional EH constraint makes the optimization more challenging to solve and the algorithms developed in  \cite{pan2019intelligent} cannot be directly applied for two reasons. Firstly, the EH constraint is non-convex. Secondly, this problem may be infeasible due to the conflicting constraints (\ref{dehwifr}) and (\ref{hhudiuwe}). In the following, we first conceive a low-complexity algorithm to solve this problem by assuming that it is feasible. Then, we study the feasibility of this problem.

\vspace{-0.2cm}\section{Low-Complexity Algorithm Development }\label{algo}

In this section, we first transform Problem (\ref{appstaoneorig}) into a more tractable one, which allows the decoupling of the TPC matrices and of the phase shifts. Then, the classic block coordinate descent (BCD) algorithm \cite{pancunhua2017} is proposed for solving the transformed problem.

\subsection{Reformulation of the Original Problem}

To deal with the complex objective function, we reformulate Problem (\ref{appstaoneorig})  by employing the well-known WMMSE method \cite{shi2011iteratively}. The appealing feature of this method is that it can transform the original complex problem into an equivalent form, which facilitates the application of the  BCD  method.

Specifically, the linear decoding matrix ${\bf{U}}$ is applied to estimate the signal vector ${{{\bf{\hat s}}}_{k}}$ for each IR
\begin{equation}\label{edwfre}
 {{{\bf{\hat s}}}_{k}} = {\bf{U}}_{k}^{\rm{H}}{{\bf{y}}_{I,k}},\forall k,
\end{equation}
where ${\bf{U}}_{k}\in \mathbb{C}^{N_I \times d}$ is the decoding matrix of the $k$th IR. Then, the MSE matrix of the $k$th IR is given by
\begin{eqnarray}
  {{\bf{E}}_{k}} &=&{\mathbb{E}_{{\bf{s,n}}}}\left[ {\left( {{{{\bf{\hat s}}}_{k}} - {{\bf{s}}_{k}}} \right){{\left( {{{{\bf{\hat s}}}_{k}} - {{\bf{s}}_{k}}} \right)}^H}} \right]\\
   &=& \!\!\left( {{\bf{U}}_{k}^{\rm{H}}{{{\bf{\bar H}}}_{k}}{{\bf{F}}_{k}} - {\bf{I}}} \right)\!\!{\left( {{\bf{U}}_{k}^{\rm{H}}{{{\bf{\bar H}}}_{k}}{{\bf{F}}_{k}}\! -\! {\bf{I}}} \right)^{\rm{H}}} +\!\!\! \sum\limits_{m = 1,m \ne k}^{K_I}\!\!\! {{\bf{U}}_{k}^{\rm{H}}{{{\bf{\bar H}}}_{k}}{{\bf{F}}_{m}}{\bf{F}}_{m}^{\rm{H}}{\bf{\bar H}}_{k}^{\rm{H}}{{\bf{U}}_{k}}} + {\sigma ^2}{\bf{U}}_{k}^{\rm{H}}{{\bf{U}}_{k}}, \forall k \in {\cal K}_I,\label{kjejofe}
\end{eqnarray}
where ${\bf{s}}$ and ${\bf{n}}$ denote the collections of data symbols and noise vectors of all IRs, respectively.

 By introducing a set of auxiliary matrices ${\bf{W}}=\{{{\bf{W}}_{k}}\succeq {\bf{0}}, \forall k \in {\cal K}_I\}$ and defining ${\bf{U}}=\{{{\bf{U}}_{k}}, \forall k \in {\cal K}_I\}$, Problem (\ref{appstaoneorig}) can be reformulated as follows \cite{pancunhua2017,shi2011iteratively}:
 \begin{subequations}\label{appstneorig}
\begin{align}
\mathop {\max }\limits_{{{\bf{W}}, {\bf{U}}, {\bf{F}},{\bm{\Phi}} } } \quad
&   \sum\limits_{k = 1}^{K_I} {{\omega_{k}}{h_{k}}\left( {{\bf{W}}, {\bf{U}},{\bf{F}},{\bm{\Phi}} } \right)}
\\
\qquad\ \textrm{s.t.}\qquad
&(\ref{dehwifr}), (\ref{hhudiuwe}), (\ref{jiofjj}),
\end{align}
\end{subequations}
where  ${h_{k}}\left( {{\bf{W}}, {\bf{U}},{\bf{F}},{\bm{\Phi}} } \right)$ is given by
\begin{equation}\label{fdgtshdy}
  {h_{k}}\left( {{\bf{W}}, {\bf{U}},{\bf{F}},{\bm{\Phi}} } \right)={\log \left| {{{\bf{W}}_{k}}} \right| - {\rm{Tr}}\left( {{{\bf{W}}_{k}}{{\bf{E}}_{k}}} \right) + d}.
\end{equation}
Although Problem (\ref{appstneorig}) has more optimization variables than Problem (\ref{appstaoneorig}), the objective function (OF) in Problem (\ref{appstneorig}) is much easier to handle, which allows the BCD algorithm to solve this problem by iteratively obtaining one set of variables while keeping the others fixed. Note that the decoding matrices ${\bf{U}}$ and the auxiliary matrices $\bf{W}$ only appear in the function ${h_{k}}\left( {{\bf{W}}, {\bf{U}},{\bf{F}}, \bm{\Phi} } \right)$. Hence, the optimal solution of ${\bf{U}}$ and $\bf{W}$ can be obtained while keeping the other matrices fixed. Specifically, given ${\bm{\Phi}}$, ${\bf{W}}$, and ${\bf{F}}$, setting the first-order derivative of ${h_{k}}\left( {{\bf{W}}, {\bf{U}},{\bf{F}}, \bm{\Phi} } \right)$ with respect to ${\bf{U}}_{k}$ and ${\bf{W}}_{k}$ to zero, we can obtain the optimal solution of
${\bf{U}}_{k}$ and ${\bf{W}}_{k}$ respectively as follows
\begin{equation}\label{fgrtgtyh}
 {\bf{U}}_{k}^\star = {\left( {{{\bf{J}}_{k}} + {{{\bf{\bar H}}}_{k}}{{\bf{F}}_{k}}{\bf{F}}_{k}^{\rm{H}}{\bf{\bar H}}_{k}^{\rm{H}}} \right)^{ - 1}}{{{\bf{\bar H}}}_{k}}{{\bf{F}}_{k}}, {\bf{W}}_{k}^\star = {\bf{E}}_{k}^{\star- 1},
\end{equation}
 where ${\bf{E}}_{k}^\star$ is obtained by inserting ${\bf{U}}_{k}^\star$ into the $k$th IR's MSE matrix in (\ref{kjejofe}), yielding
\begin{equation}\label{xsdvtg}
 {\bf{E}}_k^ \star  = {{\bf{I}}_d} - {\bf{F}}_k^{\rm{H}}{\bf{\bar H}}_k^{\rm{H}}{\left( {\sum\limits_{m = 1}^{{K_I}} {{{{\bf{\bar H}}}_k}{{\bf{F}}_m}{\bf{F}}_m^{\rm{H}}{\bf{\bar H}}_k^{\rm{H}}}  + \sigma _I^2{\bf{I}}} \right)^{ - 1}}{{{\bf{\bar H}}}_k}{{\bf{F}}_k}.
\end{equation}

In the following, we focus our attention on the optimization of TPC matrices ${\bf{F}}$  and phase shifts  ${\bm{\Phi}}$, when ${\bf{U}}$ and ${\bf{W}}$ are given.

\subsection{Optimizing the Precoding Matrices ${\bf{F}}$}\label{kodsijcosakpdc}

In this subsection, we aim to  optimize the TPC matrices ${\bf{F}}$ with fixed ${\bf{W}}, {\bf{U}}$ and $\bm{\Phi}$. By inserting ${\bf{E}}_{k}$ in (\ref{kjejofe}) into the OF of (\ref{appstneorig}) and discarding the constant terms, the TPC matrices of our optimization problem can be transformed as follows
 \begin{subequations}\label{appssxsorig}
\begin{align}
&
\mathop {\min }\limits_{{ {\bf{F}} } }\quad  \sum\limits_{k = 1}^{K_I} {{\rm{tr}}\left( {{\bf{F}}_k^{\rm{H}}{\bf{A}}{{\bf{F}}_k}} \right)} - \sum\limits_{k = 1}^{K_I} {{\omega _{k}}{\rm{Tr}}\left( {{{\bf{W}}_{k}}{\bf{U}}_{k}^{\rm{H}}{{{\bf{\bar H}}}_{k}}{{\bf{F}}_{k}}} \right)} - \sum\limits_{k = 1}^{K_I} {{\omega _{k}}{\rm{tr}}\left( {{{\bf{W}}_{k}}{\bf{F}}_{k}^{\rm{H}}{\bf{\bar H}}_{k}^{\rm{H}}{{\bf{U}}_{k}}} \right)}
\\
&\textrm{s.t.}\quad (\ref{dehwifr}), (\ref{hhudiuwe}),
\end{align}
\end{subequations}
where ${{\bf{A}} = \sum\nolimits_{m = 1}^{K_I} {{\omega _m}} {\bf{\bar H}}_m^{\rm{H}}{{\bf{U}}_m}{{\bf{W}}_m}{\bf{U}}_m^{\rm{H}}{{{\bf{\bar H}}}_m}}$.

However, due to the non-convexity of the EH constraint, Problem (\ref{appssxsorig}) is still non-convex. To resolve this issue, we observe that it can be viewed as a difference of convex (d.c.) program, which can be efficiently solved by the successive convex approximation (SCA) method \cite{cunhuajsac}. In particular, we can approximate it by its  first-order Taylor expansion. By  applying \cite[Appendix B]{pan2018robust} and Jensen' inequality, we have
\begin{equation}\label{xadeefe}
  {\rm{tr}}\left( {\sum\limits_{k = 1}^{{K_I}} {{\bf{F}}_k^{\rm{H}}{\bf{G}}{{\bf{F}}_k}} } \right)    \ge  - {\rm{tr}}\left( {\sum\limits_{k = 1}^{{K_I}} {{\bf{F}}_k^{(n){\rm{H}}}{\bf{GF}}_k^{(n)}} } \right) + 2{\mathop{\rm Re}\nolimits} \left[ {{\rm{tr}}\left( {\sum\limits_{k = 1}^{{K_I}} {{\bf{F}}_k^{(n){\rm{H}}}{\bf{G}}{{\bf{F}}_k}} } \right)} \right],
\end{equation}
where $\left\{ {{\bf{F}}_k^{(n)},\forall k} \right\}$ is the solution obtained from the previous iteration. Then, upon replacing the constraint (\ref{hhudiuwe}) by the following constraint:
\begin{equation}\label{fsfrg}
  2{\mathop{\rm Re}\nolimits} \left[ {{\rm{tr}}\left( {\sum\limits_{k = 1}^{{K_I}} {{\bf{F}}_k^{(n){\rm{H}}}{\bf{G}}{{\bf{F}}_k}} } \right)} \right] \ge \tilde Q,
\end{equation}
where $\tilde Q = \bar Q + {\rm{tr}}\left( {\sum\nolimits_{k = 1}^{{K_I}} {{\bf{F}}_k^{(n){\rm{H}}}{\bf{GF}}_k^{(n)}} } \right)$, we may  consider the following optimization problem:
 \begin{subequations}\label{apporig}
\begin{align}
&
\mathop {\min }\limits_{{ {\bf{F}} } }\quad  \sum\limits_{k = 1}^{K_I} {{\rm{tr}}\left( {{\bf{F}}_k^{\rm{H}}{\bf{A}}{{\bf{F}}_k}} \right)} - \sum\limits_{k = 1}^{K_I} {{\omega _{k}}{\rm{tr}}\left( {{{\bf{W}}_{k}}{\bf{U}}_{k}^{\rm{H}}{{{\bf{\bar H}}}_{k}}{{\bf{F}}_{k}}} \right)} - \sum\limits_{k = 1}^{K_I} {{\omega _{k}}{\rm{tr}}\left( {{{\bf{W}}_{k}}{\bf{F}}_{k}^{\rm{H}}{\bf{\bar H}}_{k}^{\rm{H}}{{\bf{U}}_{k}}} \right)}
\\
&\textrm{s.t.}\quad (\ref{dehwifr}), (\ref{fsfrg}).
\end{align}
\end{subequations}
Since the OF is convex w.r.t. $\bf{F}$, and the constraints (\ref{dehwifr}) and (\ref{fsfrg}) are convex, Problem (\ref{apporig}) constitutes a convex optimization problem, which can be solved by  standard convex solver packages, such as CVX \cite{grant2014cvx}. However, the resultant computational complexity is high. In the following, we provide a low-complexity algorithm for obtaining a nearly optimal closed-form solution by resorting to the Lagrangian dual decomposition method \cite{boyd2004convex}. Since Problem  (\ref{apporig}) is a convex problem and satisfies the slater's condition\footnote{ According to line 1 in Algorithm \ref{itedasca}, the initial precoding matrix is initialized by the solution obtained from Section \ref{frjgtoj}. Assume the original problem is feasible. Due to the randomness of channel matrices of $\bf{G}$ and $\bf{H}$, the precoding matrix obtained in Section \ref{frjgtoj} must be strictly larger than the minimum EH requirement, i.e., ${\rm{tr}}\left( {\sum\nolimits_{k = 1}^{K_I} {{\bf{F}}_k^{(0)\rm{H}}{\bf{G}}{{\bf{F}}_k^{(0)}}} } \right)> \bar Q$. Then, based on \cite{zhang2013mimo}, there must exist a strictly feasible solution, and thus the slater's condition holds. }, the dual gap is zero and the optimal solution can be obtained by solving its dual problem instead of its original one. We first introduce the Lagrange multiplier $\lambda $ associated with the power constraint, and derive the partial Lagrangian function of Problem (\ref{apporig}) as follows
\begin{equation}\label{rfgt}
  \begin{array}{l}
{\cal L}\left( {{{\bf{F}}},{\lambda}} \right) = \sum\limits_{k = 1}^{K_I} {{\rm{tr}}\left( {{\bf{F}}_{k}^{\rm{H}}{{\bf{A}}}{{\bf{F}}_{k}}} \right)}  - \sum\limits_{k = 1}^{K_I} {{\omega _{k}}{\rm{tr}}\left( {{{\bf{W}}_{k}}{\bf{U}}_{k}^{\rm{H}}{{{\bf{\bar H}}}_{k}}{{\bf{F}}_{k}}} \right)}- \sum\limits_{k = 1}^{K_I} {{\omega _{k}}{\rm{tr}}\left( {{{\bf{W}}_{k}}{\bf{F}}_{k}^{\rm{H}}{\bf{\bar H}}_{k}^{\rm{H}}{{\bf{U}}_{k}}} \right)} \\
 \qquad \qquad\qquad\quad +{\lambda}\sum\limits_{k = 1}^{K_I} {\rm{tr}}\left( {{\bf{F}}_{k}^{\rm{H}}{{\bf{F}}_{k}}} \right)  - {\lambda }{P_{T}}.
\end{array}
\end{equation}
The dual function can be obtained by solving the following problem
\begin{equation}\label{adwef}
 g\left( \lambda  \right) \buildrel \Delta \over = \mathop {\min }\limits_{\bf{F}}{\cal L}({\bf{F}},\lambda )\ \ \ \ {\rm{   s}}{\rm{.t}}{\rm{. }} \  (\ref{fsfrg}).
\end{equation}
Then, the dual problem is given by
 \begin{subequations}\label{adcsdrig}
\begin{align}
&
\mathop {\max }\limits_{{ \lambda } }\quad   g\left( \lambda  \right)
\\
&\textrm{s.t.}\quad  \lambda\ge 0.
\end{align}
\end{subequations}

Before solving the dual problem (\ref{adcsdrig}), we have to derive the expression of the dual function $g\left( \lambda  \right)$ by solving Problem (\ref{adwef}) for a given $\lambda$. By introducing the dual variable $\mu\ge 0$ associated with the constraint (\ref{fsfrg}), the Lagrangian function for Problem (\ref{adwef}) is given by
\begin{equation}\label{rdcsdewt}
  \begin{array}{l}
{\cal L}\left( {{{\bf{F}}},{\mu}} \right) = \sum\limits_{k = 1}^{K_I} {{\rm{tr}}\left( {{\bf{F}}_{k}^{\rm{H}}\left( {{{\bf{A}}} + {\lambda}{\bf{I}}} \right){{\bf{F}}_{k}}} \right)}  - \sum\limits_{k = 1}^{K_I} {{\omega _{k}}{\rm{tr}}\left( {{{\bf{W}}_{k}}{\bf{U}}_{k}^{\rm{H}}{{{\bf{\bar H}}}_{k}}{{\bf{F}}_{k}}} \right)} - \sum\limits_{k = 1}^{K_I} {{\omega _{k}}{\rm{tr}}\left( {{{\bf{W}}_{k}}{\bf{F}}_{k}^{\rm{H}}{\bf{\bar H}}_{k}^{\rm{H}}{{\bf{U}}_{k}}} \right)}\\
 \qquad \qquad\qquad\quad + \mu \tilde Q-2 \mu {\mathop{\rm Re}\nolimits} \left[ {{\rm{tr}}\left( {\sum\limits_{k = 1}^{{K_I}} {{\bf{F}}_k^{(n){\rm{H}}}{\bf{G}}{{\bf{F}}_k}} } \right)} \right]   - {\lambda }{P_{T}}.
\end{array}
\end{equation}
By setting the first-order derivative of  ${\cal L}\left( {{{\bf{F}}},{\mu}} \right)$ w.r.t. ${{{\bf{F}}_{k}^*}}$ to the zero matrix, we obtain the optimal solution of ${\bf{F}}_{k}$ as follows:
\begin{equation}\label{dwfohu}
  {{\bf{F}}_{k}^\star}(\mu) = {\left( {{\bf{A}} + \lambda {\bf{I}}} \right)^{ \dag }}\left( {{\omega _k}{\bf{\bar H}}_k^{\rm{H}}{{\bf{U}}_k}{{\bf{W}}_k} + \mu {\bf{GF}}_k^{(n)}} \right),
\end{equation}
where ${(\cdot)^\dag }$ denotes the matrix pseudoinverse. The value of $\mu$ should be chosen for ensuring that the complementary slackness condition for constraint (\ref{fsfrg}) is satisfied:
\begin{equation}\label{aswdewf}
\mu\left( 2{\mathop{\rm Re}\nolimits} \left[ {{\rm{tr}}\left( {\sum\limits_{k = 1}^{{K_I}} {{\bf{F}}_k^{(n){\rm{H}}}{\bf{G}}{{\bf{F}}_k^\star(\mu)}} } \right)} \right] - \tilde Q \right) = 0.
\end{equation}
Hence, if the following condition holds
\begin{equation}\label{cgthsth}
2{\mathop{\rm Re}\nolimits} \left[ {{\rm{tr}}\left( {\sum\limits_{k = 1}^{{K_I}} {{\bf{F}}_k^{(n){\rm{H}}}{\bf{G}}{{\bf{F}}_k^\star(0)}} } \right)} \right] \ge \tilde Q,
\end{equation}
the optimal solution of Problem (\ref{adwef}) is given by ${{\bf{F}}_{k}^\star}(0),\forall k\in {\cal K}_I$. Otherwise, the optimal $\mu$ is
\begin{equation}\label{desfreg}
\mu  = \frac{{\tilde Q - 2{\mathop{\rm Re}\nolimits} \left[ {{\rm{tr}}\left( {\sum\limits_{k = 1}^{{K_I}} {{\omega _k}{\bf{F}}_k^{(n){\rm{H}}}{\bf{G}}{{\left( {{\bf{A}} + \lambda {\bf{I}}} \right)}^{ - 1}}{\bf{\bar H}}_k^{\rm{H}}{{\bf{U}}_k}{{\bf{W}}_k}} } \right)} \right]}}{{2{\rm{tr}}\left( {\sum\limits_{k = 1}^{{K_I}} {{\bf{F}}_k^{(n){\rm{H}}}{\bf{G}}{{\left( {{\bf{A}} + \lambda {\bf{I}}} \right)}^{ - 1}}{\bf{GF}}_k^{(n)}} } \right)}}.
\end{equation}

With the aid of the dual function, we may now commence the solution of the dual problem (\ref{adcsdrig}) to find the optimal $\lambda$. Given $\lambda$, we denote the optimal solution of Problem  (\ref{adwef}) by ${{\bf{F}}_{k}}(\lambda)$. The value of $\lambda$ should be chosen for ensuring that the complementary slackness condition for the power constraint is satisfied:
\begin{equation}\label{afrgtbn}
\lambda \left( {{\rm{tr}}\left( \sum\limits_{k = 1}^{K_I} {{\bf{F}}_k^{\rm{H}}\left( \lambda  \right){{\bf{F}}_k}\left( \lambda  \right)} \right) - {P_T}} \right) = 0.
\end{equation}
If the following condition holds:
\begin{equation}\label{acsvfref}
  {{\rm{tr}}\left(\sum\limits_{k = 1}^{K_I} {{\bf{F}}_k^{\rm{H}}\left( 0  \right){{\bf{F}}_k}\left( 0  \right)} \right) \le {P_T}},
\end{equation}
then the optimal solution is given by ${\bf{F}}_k(0)$. Otherwise, we have to find  $\lambda$ for ensuring that the following equation holds:
\begin{equation}\label{qwliju}
 P(\lambda)\buildrel \Delta \over = {\rm{tr}}\left( \sum\limits_{k = 1}^{K_I} {{\bf{F}}_k^{\rm{H}}\left( \lambda  \right){{\bf{F}}_k}\left( \lambda  \right)}\right) = {P_T}.
\end{equation}
 Unfortunately, due to the complex expression of $\mu$ in  (\ref{desfreg}), we are unable to prove its monotonic nature by using the explicit expression of $P(\lambda)$ as in \cite{pan2019intelligent}.
In the following lemma, we prove that $P(\lambda)$ is a monotonically decreasing function of $\lambda$, which enables the bisection search method to find $\lambda$.

\itshape \textbf{Lemma 1:}  \upshape   The total power  $P(\lambda)$ is a monotonically decreasing function of $\lambda$.

\itshape \textbf{Proof:}  \upshape Please refer to Appendix \ref{prooflemma1}.  \hfill $\Box$

Based on Lemma 1, the bisection search method can be used for finding the solution of equation (\ref{qwliju}).  In Algorithm \ref{iteda}, we provide the detailed steps of  solving Problem (\ref{apporig}) for the case of $\lambda>0$. In each iteration of Algorithm \ref{iteda}, we have to calculate $ {{\bf{F}}_{k}^\star}(\mu) $ in (\ref{dwfohu}), which involves the calculation of ${\left( {{\bf{A}} + \lambda {\bf{I}}} \right)}^{ - 1}$ at a complexity order of ${\cal O}(N_B^3)$. If the total number of iterations is $T$, then the total complexity of calculating ${\left( {{\bf{A}} + \lambda {\bf{I}}} \right)}^{ - 1}$ is ${\cal O}(TN_B^3)$, which may be excessive. Here, we provide a method for reducing the computational complexity. Specifically, as ${{\bf{A}}}$ is a non-negative definite matrix, it can be decomposed as ${{\bf{A}}} = {\bf{Q}}{\bm{\Lambda}} {{\bf{Q}}^{\rm{H}}}$ by using the singular value decomposition (SVD), where ${\bf{Q}}{{\bf{Q}}^{\rm{H}}} = {{\bf{Q}}^{\rm{H}}}{\bf{Q}} = {{\bf{I}}_{{N_T}}}$ and ${\bm{\Lambda}}$ is a diagonal matrix with non-negative diagonal elements. Then, we have ${\left( {{\bf{A}} + \lambda {\bf{I}}} \right)}^{ - 1}={\bf{Q}}\left( {\lambda {\bf{I}} + {\bm{\Lambda}}} \right)^{-1}{{\bf{Q}}^{\rm{H}}}$. Hence, in each iteration, we only have to calculate  the product of two matrices, which has much lower complexity than calculating the inverse of the matrix having the same dimension.

 \begin{algorithm}
\caption{Bisection Search Method to Solve Problem (\ref{apporig})}\label{iteda}
\begin{algorithmic}[1]
\STATE Initialize  the accuracy $\varepsilon$, the bounds $\lambda_l$ and $\lambda_u$;
\STATE Calculate  $\lambda  = {{\left( {{\lambda _l} + {\lambda _u}} \right)} \mathord{\left/
 {\vphantom {{\left( {{\lambda _l} + {\lambda _u}} \right)} 2}} \right.
 \kern-\nulldelimiterspace} 2}$;
\STATE If condition (\ref{cgthsth}) is satisfied, $\mu$ is equal to zero. Otherwise, update $\mu$ in (\ref{desfreg});
\STATE Calculate $\{{\bf{F}}_{k}(\lambda), \forall k \}$ according to (\ref{dwfohu});
 \STATE  If $P(\lambda)\ge P_T$, set ${\lambda _l}={\lambda}$. Otherwise, set ${\lambda _u}={\lambda}$;
 \STATE  If  $\left| {{\lambda _l} - {\lambda _u}} \right| \le \varepsilon $, terminate. Otherwise, go to step 2.
\end{algorithmic}
\end{algorithm}

Based on the above discussions, in Algorithm \ref{itedasca} we provide the detailed steps of the SCA algorithm conceived for solving Problem (\ref{appssxsorig}).

 \begin{algorithm}
\caption{SCA Algorithm to Solve Problem (\ref{appssxsorig})}\label{itedasca}
\begin{algorithmic}[1]
\STATE Initialize  the accuracy $\varepsilon$, the precoding matrices ${\bf{F}}^{(0)}$ from Section \ref{itedasca}, the iteration index $n=0$, the maximum number of iterations $n_{\rm{max}}$, calculate the OF value of Problem (\ref{appssxsorig}) as $z({\bf{F}}^{(0)})$;
\STATE Calculate  $\tilde Q^{(n)} = \bar Q + {\rm{tr}}\left( {\sum\nolimits_{k = 1}^{{K_I}} {{\bf{F}}_k^{(n){\rm{H}}}{\bf{GF}}_k^{(n)}} } \right)$;
\STATE With $\tilde Q^{(n)} $, calculate $\{{\bf{F}}_{k}^{(n+1)}, \forall k \}$ by solving Problem (\ref{apporig}) using Algorithm \ref{iteda};
\STATE  If $n\geq n_{\rm{max}}$ or ${{\left| {z({{\bf{F}}^{(n + 1)}}) - z({{\bf{F}}^{(n)}})} \right|} \mathord{\left/
 {\vphantom {{\left| {z({{\bf{F}}^{(n + 1)}}) - z({{\bf{F}}^{(n)}})} \right|} {\left| {z({{\bf{F}}^{(n+1)}})} \right|}}} \right.
 \kern-\nulldelimiterspace} {\left| {z({{\bf{F}}^{(n+1)}})} \right|}} < \varepsilon $, terminate.  Otherwise, set $n \leftarrow n + 1$  and go to step 2.
\end{algorithmic}
\end{algorithm}

In the following, we show that Algorithm \ref{itedasca} converges to the KKT point of Problem (\ref{appssxsorig}).

\itshape \textbf{Theorem 1:}  \upshape  The sequences of $\{{\bf{F}}^{(n)}, n=1,2,\cdots\}$ generated by Algorithm \ref{itedasca}  converge to the KKT optimum point of Problem (\ref{appssxsorig}).

\itshape \textbf{Proof:}  \upshape The proof is similar to that of \cite{cunhuawcl} and  hence it is omitted for simplicity.  \hfill $\Box$

 Next, we briefly analyze the complexity of Algorithm \ref{itedasca}. We assume that $N_B\ge N_I\ge d$. In each iteration of Algorithm \ref{itedasca}, the main complexity contribution is the calculation of $\{{\bf{F}}_{k}^{(n+1)}, \forall k \}$ by using the bisection search method in Algorithm \ref{iteda}. In each iteration of Algorithm \ref{iteda}, the main complexity lies in calculating $\bf{F}$ in (\ref{dwfohu}),  which is on the order of ${\cal O}(K_IN_B^3)$. The number of iterations required for Algorithm \ref{iteda} to converge is given by ${\log _2}\left( {\frac{{{\lambda _u} - {\lambda _l}}}{\varepsilon }} \right)$. Hence, the total complexity of Algorithm \ref{iteda} is ${\cal O}({\log _2}\left( {\frac{{{\lambda _u} - {\lambda _l}}}{\varepsilon }} \right) K_IN_B^3)$. Then, the total complexity of Algorithm \ref{itedasca} is given by ${\cal O}(n_{\rm{max}}{\log _2}\left( {\frac{{{\lambda _u} - {\lambda _l}}}{\varepsilon }} \right) K_IN_B^3)$.

\subsection{Optimizing the Phase Shift Matrix} 

In this subsection, we focus our attention on optimizing the phase shift matrix ${\bm{\Phi}}$, while fixing the other parameters. Upon substituting ${\bf{E}}_{k}$ in (\ref{kjejofe}) into (\ref{fdgtshdy}) and removing the terms that are independent of ${\bm{\Phi}}$, the phase shift optimization problem is formulated as:
 \begin{subequations}\label{appsdsworig}
\begin{align}
\!&{\mathop {\min }\limits_{{\bm{\Phi}}} \quad \!\!\!\! \sum\limits_{k = 1}^{{K_I}} {{\rm{tr}}\left(\! {{\omega _k}{{\bf{W}}_k}{\bf{U}}_k^{\rm{H}}{{{\bf{\bar H}}}_k}{{\bf{\tilde F}}}{\bf{\bar H}}_k^{\rm{H}}{{\bf{U}}_k}} \!\right)}\!\!-\!   {\sum\limits_{k = 1}^{K_I} {{\rm{tr}}\left( {\omega _{k}}{{{\bf{W}}_{k}}{\bf{U}}_{k}^{\rm{H}}{{{\bf{\bar H}}}_{k}}{{\bf{F}}_{k}}} \right)} }\!-\!{\sum\limits_{k = 1}^{K_I} {{\rm{tr}}\left({\omega _{k}} {{{\bf{W}}_{k}}{\bf{F}}_{k}^{\rm{H}}{\bf{\bar H}}_{k}^{\rm{H}}{{\bf{U}}_{k}}} \right)} }   }\label{odjfsoer}
\\
\!&\textrm{s.t.}\quad (\ref{hhudiuwe}), (\ref{jiofjj}),
\end{align}
\end{subequations}
where ${\bf{\tilde F}}=\sum\nolimits_{m = 1}^{K_I}{{\bf{F}}_{m}}{\bf{F}}_{m}^{\rm{H}}$.

By substituting  $ {\bf{\bar H}}_{k}  =  {{{\bf{H}}_{b,k}} + {{\bf{H}}_{r,k}}{\bf{\Phi}} {\bf{Z}}}$ into (\ref{odjfsoer}), we have
\begin{equation}\label{adsqa}
\begin{array}{l}
\!\!{\omega _{k}}{{\bf{W}}_{k}}{\bf{U}}_{k}^{\rm{H}}{{{\bf{\bar H}}}_{k}}{{\bf{\tilde F}} }{\bf{\bar H}}_{k}^{\rm{H}}{{\bf{U}}_{k}} \!=\! {\omega _{k}}{{\bf{W}}_{k}}{\bf{U}}_{k}^{\rm{H}}{{\bf{H}}_{r,k}}{\bm{\Phi}} {{\bf{Z}} }{{\bf{\tilde F}} }{\bf{Z}}^{\rm{H}}{{\bm{\Phi}} ^{\rm{H}}}{\bf{H}}_{r,k}^{\rm{H}}{{\bf{U}}_{k}} \!+\! {\omega _{k}} {{\bf{W}}_{k}}{\bf{U}}_{k}^{\rm{H}}{{\bf{H}}_{b,k}}{{\bf{\tilde F}} }{\bf{Z}}^{\rm{H}}{{\bm{\Phi}} ^{\rm{H}}}{\bf{H}}_{r,k}^{\rm{H}}{{\bf{U}}_{k}}\\
\qquad\qquad\quad \qquad\qquad\quad +{\omega _{k}} {{\bf{W}}_{k}}{\bf{U}}_{k}^{\rm{H}}{{\bf{H}}_{r,k}}{\bm{\Phi}} {{\bf{Z}}}{{\bf{\tilde F}}}{\bf{H}}_{b,k}^{\rm{H}}{{\bf{U}}_{k}} + {\omega _{k}} {{\bf{W}}_{k}}{\bf{U}}_{k}^{\rm{H}}{{\bf{H}}_{b,k}}{{\bf{\tilde F}} }{\bf{H}}_{b,k}^{\rm{H}}{{\bf{U}}_{k}},
\end{array}
\end{equation}
and
\begin{equation}\label{dwgtrh}
 {\omega _{k}}{{\bf{W}}_{k}}{\bf{U}}_{k}^{\rm{H}}{{{\bf{\bar H}}}_{k}}{{\bf{F}}_{k}} = {\omega _{k}} {{\bf{W}}_{k}}{\bf{U}}_{k}^{\rm{H}}{{\bf{H}}_{r,k}}{\bm{\Phi}}{{\bf{Z}} }{{\bf{F}}_{k}} + {\omega _{k}} {{\bf{W}}_{k}}{\bf{U}}_{k}^{\rm{H}}{{\bf{H}}_{b,k}}{{\bf{F}}_{k}}.
\end{equation}

Let us define
${{\bf{B}}_{k}} \buildrel \Delta \over =  {\omega _{k}}{\bf{H}}_{r,k}^{\rm{H}}{{\bf{U}}_{k}}{{\bf{W}}_{k}}{\bf{U}}_{k}^{\rm{H}}{{\bf{H}}_{r,k}}$, ${{\bf{C}} } \buildrel \Delta \over =  {{\bf{Z}}}{{\bf{\tilde F}}}{\bf{Z}}^{\rm{H}}$ and ${{\bf{D}}_{ k}} \buildrel \Delta \over = {\omega _{k}}{{\bf{Z}} }{\bf{\tilde F}}^{\rm{H}}{\bf{H}}_{b,k}^{\rm{H}}{{\bf{U}}_{k}}{{\bf{W}}_{k}}{\bf{U}}_{k}^{\rm{H}}{{\bf{H}}_{r,k}}$. By using (\ref{adsqa}), we arrive at:
\begin{equation}\label{wdfwfgr}
  {\rm{tr}}\left({\omega _{k}} {{{\bf{W}}_{k}}{\bf{U}}_{k}^{\rm{H}}{{{\bf{\bar H}}}_{k}}{{\bf{\tilde F}}}{\bf{\bar H}}_{k}^{\rm{H}}{{\bf{U}}_{k}}}\right)= {\rm{tr}}\left( {{{\bm{\Phi}} ^{\rm{H}}}{{\bf{B}}_{k}}{\bm{\Phi}} {{\bf{C}}}} \right) + {\rm{tr}}\left( {{{\bm{\Phi}} ^{\rm{H}}}{{\bf{D}}_{k}^{\rm{H}}}} \right) +   {\rm{tr}}\left( {{\bm{\Phi}} {\bf{D}}_{k} } \right) + {\rm{const}_1},
\end{equation}
where ${\rm{const}_1}$ is a constant term that is independent of ${\bm{\Phi}}$.

Similarly, by defining  ${{\bf{T}}_{k}} \buildrel \Delta \over = {\omega _{k}}{{\bf{Z}}}{{\bf{F}}_{k}}{{\bf{W}}_{k}}{\bf{U}}_{k}^{\rm{H}}{{\bf{H}}_{r,k}}$, from (\ref{dwgtrh}) we have
\begin{equation}\label{edwf}
 {\rm{tr}}\left( {{\omega _{k}}{{\bf{W}}_{k}}{\bf{U}}_{k}^{\rm{H}}{{{\bf{\bar H}}}_{k}}{{\bf{F}}_{k}}} \right) = {\rm{tr}}\left({\bm{\Phi}} {{\bf{T}}_{k}}\right) + {\rm{const}}_2,
\end{equation}
where ${\rm{const}}_2$ is a constant term that is independent of ${\bm{\Phi}}$.

By defining ${{\bf{G}}_b} \buildrel \Delta \over = \sum\nolimits_{l = 1}^{{K_E}} {{\alpha _l}\eta } {\bf{G}}_{b,l}^{\rm{H}}{{\bf{G}}_{b,l}}$, ${{\bf{G}}_r} \buildrel \Delta \over = \sum\nolimits_{l = 1}^{{K_E}} {{\alpha _l}\eta } {\bf{G}}_{r,l}^{\rm{H}}{{\bf{G}}_{r,l}}$, and ${{\bf{G}}_{br}} \buildrel \Delta \over = {\bf{Z\tilde F}}\sum\nolimits_{l = 1}^{{K_E}} {{\alpha _l}\eta } {\bf{G}}_{b,l}^{\rm{H}}{{\bf{G}}_{r,l}}$,
the EH constraint in (\ref{hhudiuwe}) can be recast as follows:
\begin{equation}\label{acvfeatgb}
  {\rm{tr}}\left( {{{\bm{\Phi}} ^{\rm{H}}}{{\bf{G}}_r}{\bm{\Phi}}{\bf{C}}} \right) + {\rm{tr}}\left( {{{\bm{\Phi}} ^{\rm{H}}}{\bf{G}}_{b,r}^{\rm{H}}} \right) + {\rm{tr}}\left( {{\bm{\Phi}} {{\bf{G}}_{br}}} \right) + {\rm{tr}}\left( {{{\bf{G}}_b}{\bf{\tilde F}}} \right) \ge \bar Q.
\end{equation}

By inserting (\ref{wdfwfgr}) and (\ref{edwf}) into the OF of Problem (\ref{appsdsworig}) and removing the constant terms,  we have
 \begin{subequations}\label{appjswifhorig}
\begin{align}
&{\mathop {\min }\limits_{{\bm{\Phi}}}  \quad {\rm{tr}}\left( {{{\bm{\Phi}} ^{\rm{H}}}{\bf{B}}{\bm{\Phi}} {\bf{C}}} \right) + {\rm{tr}}\left( {{{\bm{\Phi}} ^{\rm{H}}}{{\bf{V}}^{\rm{H}}}} \right) + {\rm{tr}}\left( {{\bm{\Phi}} {\bf{V}}} \right)}
\\
&\textrm{s.t.}\quad (\ref{jiofjj}), (\ref{acvfeatgb}),
\end{align}
\end{subequations}
where ${\bf{B}}$ and ${\bf{V}}$ are  given by ${\bf{B}} =   {\sum\nolimits_{k = 1}^{K_I} {{{\bf{B}}_{k}}} } $ and ${\bf{V}} =     \sum\nolimits_{k = 1}^{K_I} {{{\bf{D}}_{k}}}      -   \sum\nolimits_{k = 1}^{K_I} {{{\bf{T}}_{ k}}}$, respectively.

Upon denoting  the collection of diagonal elements of ${\bm{\Phi}}$ by ${\bm{\phi}}={\left[ {{\phi _1}, \cdots ,{\phi _M}} \right]^{\rm{T}}}$ and adopting the matrix identity of \cite[Eq. (1.10.6)]{zhang2017matrix}, it follows that
\begin{equation}\label{saddewde}
  {\rm{tr}}\left( {{{\bm{\Phi}} ^{\rm{H}}}{\bf{B}}{\bm{\Phi}} {\bf{C}}} \right) = {{\bm{\phi}} ^{\rm{H}}}\left( {{\bf{B}} \odot {{\bf{C}}^{\rm{T}}}} \right){\bm{\phi}}, {\rm{tr}}\left( {{{\bm{\Phi}} ^{\rm{H}}}{{\bf{G}}_r}{\bm{\Phi}}{\bf{C}}} \right)= {{\bm{\phi}} ^{\rm{H}}}\left( {{\bf{G}}_r \odot {{\bf{C}}^{\rm{T}}}} \right){\bm{\phi}}.
\end{equation}
Upon denoting the collections of diagonal elements of ${\bf{V}}$ and ${\bf{G}}_{br}$ by ${\bf{v}} = {\left[ {{{\left[ {\bf{V}} \right]}_{1,1}}, \cdots ,{{\left[ {\bf{V}} \right]}_{M,M}}} \right]^{\rm{T}}}$ and ${\bf{g}} = {\left[ {{{\left[ {\bf{G}}_{br}\right]}_{1,1}}, \cdots ,{{\left[ {\bf{G}}_{br} \right]}_{M,M}}} \right]^{\rm{T}}}$, we arrive at
\begin{equation}\label{sdewf}
  {\rm{tr}}\left( {{\bm{\Phi}} {\bf{V}}} \right)={{\bf{v}}^{\rm{T}}}{{\bm{\phi}} }, {\rm{tr}}\left( {{{\bm{\Phi}} ^{\rm{H}}}{{\bf{V}}^{\rm{H}}}} \right) = {\bm{\phi}}^{\rm{H}}{\bf{v}^*}, {\rm{tr}}\left( {{\bm{\Phi}} {{\bf{G}}_{br}}} \right)= {{\bf{g}}^{\rm{T}}}{{\bm{\phi}}}, {\rm{tr}}\left( {{{\bm{\Phi}} ^{\rm{H}}}{{{\bf{G}}_{br}^{\rm{H}}}}} \right) = {\bm{\phi}}^{\rm{H}}{\bf{g}^*}.
\end{equation}
Moreover, the constraint (\ref{acvfeatgb}) can be rewritten as
\begin{equation}\label{acdfwefe}
  {{\bm{\phi}} ^{\rm{H}}}\bm{\Upsilon}{\bm{\phi}}+2{\mathop{\rm Re}\nolimits} \left\{ {{{\bm{\phi}} ^{\rm{H}}}{\bf{g}^*}} \right\} \ge {\mathord{\buildrel{\lower3pt\hbox{$\scriptscriptstyle\frown$}}
\over Q} },
\end{equation}
where we have ${\mathord{\buildrel{\lower3pt\hbox{$\scriptscriptstyle\frown$}}
\over Q} }= \bar Q-{\rm{Tr}}\left( {{{\bf{G}}_b}{\bf{\tilde F}}} \right)$ and $\bm{\Upsilon}={{\bf{G}}_r \odot {{\bf{C}}^{\rm{T}}}} $. It can be   verified that ${\bf{G}}_r$ and ${{\bf{C}}^{\rm{T}}}$ are non-negative semidefinite matrices. Then, according to   \cite{zhang2017matrix}, the Hadamard product ${{\bf{G}}_r\odot {{\bf{C}}^{\rm{T}}}}$ (or equivalently $\bm{\Upsilon}$) is also a semidefinite matrix.

Thus, Problem (\ref{appjswifhorig}) can be transformed as
 \begin{subequations}\label{appjhorig}
\begin{align}
&{\mathop {\min }\limits_{\bm{\phi}}  \quad {{\bm{\phi}} ^{\rm{H}}}{\bm{\Xi} }{\bm{\phi}} + 2{\mathop{\rm Re}\nolimits} \left\{ {{{\bm{\phi}} ^{\rm{H}}}{\bf{v}^*}} \right\}  }
\\
&\textrm{s.t.}\quad (\ref{jiofjj}), (\ref{acdfwefe}),
\end{align}
\end{subequations}
where we have $\bm{\Xi}={{\bf{B}} \odot {{\bf{C}}^{\rm{T}}}} $.  Again, ${\bf{B}}$  can be verified to be a non-negative semidefinite matrix, and thus ${\bm{\Xi} } $ is a non-negative semidefinite matrix.

Due to the non-convex constraint (\ref{acdfwefe}), Problem (\ref{appjhorig}) is difficult to solve. To deal with this constraint, we again employ the SCA method \cite{cunhuajsac}.  Specifically, since ${{\bm{\phi}} ^{\rm{H}}}\bm{\Upsilon}{\bm{\phi}}$ is convex w.r.t. ${\bm{\phi}}$, its lower bound can be obtained as follows:
\begin{equation}\label{xaxasxaefe}
  {{\bm{\phi}} ^{\rm{H}}}\bm{\Upsilon}{\bm{\phi}} \ge  - {{\bm{\phi}} ^{{(n)}\rm{H}}}\bm{\Upsilon}{\bm{\phi}^{(n)}} + 2{\mathop{\rm Re}\nolimits} \left[  {{\bm{\phi}} ^{\rm{H}}}\bm{\Upsilon}{\bm{\phi}^{(n)}} \right],
\end{equation}
where ${{\bm{\phi}} ^{{(n)}}}$ is obtained in the previous iteration.
Then, constraint (\ref{acdfwefe}) is replaced by the following constraint
\begin{equation}\label{sacvfgatr}
 2{\mathop{\rm Re}\nolimits} \left[ {{{\bm{\phi}} ^{\rm{H}}}\left( {{\bf{g}^*} + {\bm{\Upsilon} }{{\bm{\phi}} ^{(n) }}} \right)} \right] \ge {\mathord{\buildrel{\lower3pt\hbox{$\scriptscriptstyle\frown$}}
\over Q}  + {{\bm{\phi}} ^{(n){\rm{H}}}}\bm{\Upsilon} {{\bm{\phi}} ^{(n)}}}\buildrel \Delta \over = \hat Q,
\end{equation}
which is a linear constraint. Then, Problem (\ref{appjhorig}) then becomes
 \begin{subequations}\label{apdwedg}
\begin{align}
&{\mathop {\min }\limits_{\bm{\phi}}  \quad {{\bm{\phi}} ^{\rm{H}}}{\bm{\Xi} }{\bm{\phi}} +2{\mathop{\rm Re}\nolimits} \left\{ {{{\bm{\phi}} ^{\rm{H}}}{\bf{v}^*}} \right\} }
\\
&\textrm{s.t.}\quad (\ref{jiofjj}), (\ref{sacvfgatr}).
\end{align}
\end{subequations}

In the following, we conceive the Majorization-Minimization (MM) algorithm \cite{yansun} for  solving Problem (\ref{apdwedg}). The key idea is to solve a challenging problem by introducing a series of more tractable subproblems. Upon denoting the objective function of Problem (\ref{apdwedg}) by $f({\bm{\phi}})$, in the $(n+1)$th iteration we have to find the upper bound of the OF, denoted as $g({\bm{\phi}}|{\bm{\phi}}^{(n)})$, which should satisfy the following three conditions:
\begin{equation}\label{adefe}
  1) g({\bm{\phi}}^{(n)}|{\bm{\phi}}^{(n)})\!=\!f({\bm{\phi}}^{(n)}); 2) {\left. {{\nabla _{\bm{\phi}^*} }g({\bm{\phi}} |{{\bm{\phi}} ^{(n)}})} \right|_{{\bm{\phi}}  = {{\bm{\phi}} ^{(n)}}}} = {\left. {{\nabla _{\bm{\phi}^*} }f({\bm{\phi}} )} \right|_{{\bm{\phi}}  = {{\bm{\phi}}^{(n)}}}}; 3) g({\bm{\phi}}|{\bm{\phi}}^{(n)})\!\ge\! f({\bm{\phi}}).
\end{equation}
Then, we solve the approximate subproblem defined by a more tractable new OF $g({\bm{\phi}}|{\bm{\phi}}^{(n)})$. To find $g({\bm{\phi}}|{\bm{\phi}}^{(n)})$, we introduce the following lemma \cite{jiansong}.

\itshape \textbf{Lemma 2:}  \upshape  For any given  ${\bm{\phi}}^{(n)}$, the following inequality holds for any feasible ${\bm{\phi}}$:
\begin{equation}\label{sxdwef}
  {{\bm{\phi}} ^{\rm{H}}}{\bm{\Xi}}{\bm{\phi}}  \le {{\bm{\phi}} ^{\rm{H}}}{\bf{X}}{\bm{\phi}}  - 2{\mathop{\rm Re}\nolimits} \left\{ {{{\bm{\phi}} ^{\rm{H}}}\left( {{\bf{X}} - {\bm{\Xi}}} \right){{\bm{\phi}} ^{(n)}}} \right\} + {\left( {{{\bm{\phi}} ^{(n)}}} \right)^{\rm{H}}}\left( {{\bf{X}} - {\bm{\Xi}}} \right){{\bm{\phi}}^{(n)}} \buildrel \Delta \over = y({\bm{\phi}} |{{\bm{\phi}}^{(n)}}),
\end{equation}
where ${\bf{X}} = {\lambda _{\rm{\max} }}{{\bf{I}}_M}$ and ${\lambda _{\rm{\max} }}$ is the maximum eigenvalue of ${\bm{\Xi}}$.  \hfill $\Box$

Then, the function $g({\bm{\phi}}|{\bm{\phi}}^{(n)})$  can be constructed as follows:
\begin{equation}\label{sdwxs}
 g({\bm{\phi}}|{\bm{\phi}}^{(n)})=y({\bm{\phi}} |{{\bm{\phi}}^{(n)}})+ 2{\rm{Re}}\left\{ {{{\bm{\phi}} ^{\rm{H}}}{{\bf{v}}^*}} \right\},
\end{equation}
where $y({\bm{\phi}} |{{\bm{\phi}}^{(n)}})$ is defined in (\ref{sxdwef}).   The new OF $g({\bm{\phi}}|{\bm{\phi}}^{(n)})$ is more tractable  than the original OF $f({\bm{\phi}})$. The subproblem to be solved is given by
 \begin{subequations}\label{apsasdcjig}
\begin{align}
&{\mathop {\min }\limits_{\bm{\phi}}  \quad g({\bm{\phi}}|{\bm{\phi}}^{(n)})}
\\
&\textrm{s.t.}\quad (\ref{jiofjj}), (\ref{sacvfgatr}).
\end{align}
\end{subequations}
Since ${\bm{\phi}} ^{\rm{H}}{\bm{\phi}}=M$, we have  ${{\bm{\phi}} ^{\rm{H}}}{\bf{X}}{\bm{\phi}}=M\lambda _{\rm{\max} }$, which is a constant. By removing the other constants, Problem (\ref{apsasdcjig}) can be rewritten as follows:
 \begin{subequations}\label{asasdcjig}
\begin{align}
&{\mathop {\max }\limits_{\bm{\phi}}  \quad 2{\mathop{\rm Re}\nolimits} \left\{ {{{\bm{\phi}} ^{\rm{H}}{\bf{q}}^{(n)}} } \right\}}
\\
&\textrm{s.t.}\quad (\ref{jiofjj}), (\ref{sacvfgatr}),
\end{align}
\end{subequations}
where ${\bf{q}}^{(n)}=\left( {{\lambda _{\rm{\max} }}{{\bf{I}}_M} - {\bm{\Xi}}} \right){{\bm{\phi}} ^{(n)}}-{\bf{v}}^*.$ Due to the additional constraint (\ref{sacvfgatr}), the optimal solution of Problem (\ref{asasdcjig}) cannot be  obtained as in \cite{pan2019intelligent}. Furthermore, due to the non-convex unit-modulus constraint (\ref{jiofjj}), Problem (\ref{asasdcjig}) is a non-convex optimization problem. As a result, the Lagrangian dual decomposition method developed for the convex problem (\ref{apporig}) is not applicable here, since the dual gap is not zero.

 In the following, we propose a price mechanism for  solving Problem (\ref{asasdcjig}) that can obtain the globally optimal solution.   Specifically, we consider the following problem by introducing a non-negative price  $p$ on the left hand side of constraint (\ref{sacvfgatr}):
 \begin{subequations}\label{asasxwdwcjig}
\begin{align}
&{\mathop {\max }\limits_{\bm{\phi}}  \quad 2{\mathop{\rm Re}\nolimits} \left\{ {{{\bm{\phi}} ^{\rm{H}}{\bf{q}}^{(n)}} } \right\}}+2p{\mathop{\rm Re}\nolimits} \left[ {{{\bm{\phi}} ^{\rm{H}}}\left( {{\bf{g}^*} + {\bm{\Upsilon} }{{\bm{\phi}} ^{(n) }}} \right)} \right]
\\
&\textrm{s.t.}\quad (\ref{jiofjj}).
\end{align}
\end{subequations}
For a given $p$, the globally optimal solution is given by
\begin{equation}\label{asxwed}
  \bm{\phi}{(p)}={e^{j\arg \left({\bf{q}}^{(n)}+p\left( {{\bf{g}^*} + {\bm{\Upsilon} }{{\bm{\phi}} ^{(n) }}} \right)\right)}}.
\end{equation}
Our objective is to find a $p$ value for ensuring that the complementary slackness condition for constraint (\ref{sacvfgatr}) is satisfied:
\begin{equation}\label{wdefr}
  p \left( J(p)-\hat Q\right)=0,
\end{equation}
where $J(p)=2{\mathop{\rm Re}\nolimits} \left[ {{{\bm{\phi}}{(p)} ^{\rm{H}}}\left( {{\bf{g}^*} + {\bm{\Upsilon} }{{\bm{\phi}} ^{(n) }}} \right)} \right]$.
To solve this equation, we consider two cases: 1) $p=0$; 2) $p>0$.

\emph{\textbf{Case I:}} In this case, $\bm{\phi}{(0)}={e^{j\arg \left({\bf{q}}^{(n)}\right)}}$ has to satisfy constraint (\ref{sacvfgatr}). Otherwise, $p>0$.

\emph{\textbf{Case II:}}   Since $p>0$, equation (\ref{wdefr}) holds only when $J(p) =\hat Q$. To solve this equation, we first provide the following lemma.

\itshape \textbf{Lemma 3:}  \upshape   Function  $J(p)$ is a monotonically increasing function of $p$.

\itshape \textbf{Proof:}  \upshape The proof is similar to Lemma 1 and thus omitted.  \hfill $\Box$

Based on Lemma 3, the bisection search method can be adopted for finding the  solution of $J(p) =\hat Q$. Based on the above discussions, we provide the algorithm to solve Problem (\ref{asasdcjig}) in Algorithm \ref{isaaada}.
Although Problem (\ref{asasdcjig}) is a non-convex problem, in the following theorem we prove that Algorithm \ref{isaaada} is capable of finding the globally optimal solution.

\itshape \textbf{Theorem 2:}  \upshape   Algorithm \ref{isaaada} is capable of finding the globally optimal solution of Problem (\ref{asasdcjig}) and thus also of Problem (\ref{apsasdcjig}).

\itshape \textbf{Proof:}  \upshape Please refer to Appendix \ref{prooftheorem2}.  \hfill $\Box$

 \begin{algorithm}
\caption{Bisection Search Method to Solve Problem (\ref{asasdcjig})}\label{isaaada}
\begin{algorithmic}[1]
\STATE Calculate $J(0)$. If $J(0)\le \hat Q$, terminate. Otherwise, go to step 2.
\STATE Initialize  the accuracy $\varepsilon$, bounds $p_l$ and $p_u$;
\STATE Calculate  $p = {{\left( {{p_l} + {p_u}} \right)} \mathord{\left/
 {\vphantom {{\left( {{p_l} + {p_u}} \right)} 2}} \right.
 \kern-\nulldelimiterspace} 2}$;
\STATE Update $\bm{\phi}{(p)}$ in (\ref{asxwed}) and calculate $J(p)$;
 \STATE  If $J(p) \ge \hat Q$, set ${p _u}=p$; Otherwise, set ${p _l}=p$;
 \STATE  If  $\left| {{p _l} - {p _u}} \right| \le \varepsilon $, terminate; Otherwise, go to step 3.
\end{algorithmic}
\end{algorithm}

Based on the above, we now provide the details of solving Problem (\ref{appsdsworig}) in Algorithm \ref{csdcaada}.

 \begin{algorithm}
\caption{MM Combined with SCA Algorithm to Solve Problem (\ref{appsdsworig})}\label{csdcaada}
\begin{algorithmic}[1]
\STATE Initialize  the accuracy $\varepsilon$,  the phase shifts ${\bm{\phi}}^{(0)}$, the iteration index to $n=0$, the maximum number of iterations to  $n_{\rm{max}}$, calculate the OF value of Problem (\ref{apdwedg}) as $f({\bm{\phi}}^{(0)})$;
\STATE Calculate $\hat Q^{(n)}={\mathord{\buildrel{\lower3pt\hbox{$\scriptscriptstyle\frown$}}
\over Q}  + {{\bm{\phi}} ^{(n){\rm{H}}}}\bm{\Upsilon} {{\bm{\phi}} ^{(n)}}} $;
\STATE Calculate ${\bf{q}}^{(n)}=\left( {{\lambda _{\rm{\max} }}{{\bf{I}}_M} - {\bm{\Xi}}} \right){{\bm{\phi}} ^{(n)}}-{\bf{v}}^*$;
\STATE Update ${{\bm{\phi}}^{(n+1)}} $ by solving Problem (\ref{asasdcjig}) using Algorithm \ref{isaaada};
 \STATE  If $n\geq n_{\rm{max}}$ or ${{\left| {f({\bm{\phi} ^{(n+1)}}) - f({\bm{\phi} ^{(n)}})} \right|} \mathord{\left/
 {\vphantom {{\left| {f({\bm{\phi} ^{(n+1)}}) - f({\bm{\phi} ^{(n) }})} \right|} {f({\bm{\phi} ^{(n+1)}})}}} \right.
 \kern-\nulldelimiterspace} {f({\bm{\phi} ^{(n+1)}})}} \le \varepsilon $ holds, terminate; Otherwise, set $n \leftarrow n + 1$ and go to step 2.
\end{algorithmic}
\end{algorithm}

In the following theorem, we prove that the sequence of $\{{\bm{\phi} ^{(n)}},n=1,2,\cdots\}$ generated by Algorithm
\ref{csdcaada} converges to the KKT-optimal point of Problem (\ref{appsdsworig}).

\itshape \textbf{Theorem 3:}  \upshape  The sequences of the OF value produced by Algorithm \ref{csdcaada} are guaranteed to converge, and the final solution satisfies the KKT point of Problem  (\ref{appsdsworig}).

\itshape \textbf{Proof:}  \upshape Please refer to Appendix \ref{prooftheorem3}.  \hfill $\Box$

 Let us now analyze the complexity of Algorithm \ref{csdcaada}. The complexity is dominated by calculating  ${{\bm{\phi}}^{(n+1)}} $ in step 4 using Algorithm \ref{isaaada}. The complexity   mainly depends on calculating the maximum eigenvalue of ${\bm{\Xi}}$. Its complexity is on the order of ${\cal O}(M^3)$. The number of iterations required for Algorithm  \ref{isaaada} is  ${\log _2}\left( {\frac{{{p _u} - {p _l}}}{\varepsilon }} \right)$. Then, the total complexity of step 3 is ${\cal O}({\log _2}\left( {\frac{{{p _u} - {p _l}}}{\varepsilon }} \right) M^3)$. Hence, the total complexity of Algorithm \ref{csdcaada} is given by ${\cal O}(n_{\rm{max}}{\log _2}\left( {\frac{{{p _u} - {p _l}}}{\varepsilon }} \right) M^3)$.

%\subsubsection{Complexity Analysis}
%
%In this part, we analyze the complexity of the proposed two methods to solve Problem (\ref{appjig}).
%
%
%We now analyze the complexity of MM algorithm. At the beginning of the MM algorithm, one needs to calculate ${\lambda _{\rm{\max} }}$, i.e., the maximum eigenvalue of ${\bm{\Xi}}$. The complexity is given by ${\cal O}(M^3)$. For each iteration of the MM algorithm, the main complexity lies in the calculation of ${\bf{q}}^t$ in step 2, the complexity of which is ${\cal O}(M^2)$. Denote $T_{MM}$ as the number of iterations required for the MM algorithm to converge. Then, the total complexity of the MM algorithm is given by $C_{\rm{MM}}={\cal O}(M^3+T_{MM}M^2)$.
%
%We then analyze the complexity of the CCM algorithm. At the start of the CCM algorithm, we need to find the range of
%$\alpha $ and $\beta$ to guarantee the convergence of the CCM algorithm, which needs to calculate the largest eigenvalue of matrices $\bm{\Xi}$ ($\lambda _{\bm{\Xi}}$) as shown in Theorem 1. Its complexity is given by ${\cal O}(M^3)$. For each iteration of the CCM algorithm, the complexity  mainly depends on the calculation of Euclidean gradient ${\bm{\eta} ^t} $, which is given by ${\cal O}(M^2)$. Denote $T_{CCM}$ as the total number of iterations for the CCM algorithm to converge. The total complexity of the CCM algorithm is given by $C_{\rm{CCM}}={\cal O}(M^3+T_{CCM}M^2)$.

\subsection{Overall Algorithm to Solve Problem (\ref{appstaoneorig})}

Based on the above analysis, we provide the detailed steps of the BCD algorithm to solve Problem (\ref{appstaoneorig}) in Algorithm \ref{bcd}, where $R({\bf{F}}^{(n)},{\bm{\phi}}^{(n)})$ denotes the OF value of Problem (\ref{appstaoneorig}) in the $n$th iteration.

\begin{algorithm}
\caption{Block Coordinate Descent Algorithm}\label{bcd}
\begin{algorithmic}[1]
\STATE Initialize iterative number $n=1$, maximum number of iterations $n_{\rm{max}}$,  feasible ${\bf{F}}^{(1)}$,
${\bm{\phi}}^{(1)}$, error tolerance $\varepsilon$, calculate $R({\bf{F}}^{(1)},{\bm{\phi}}^{(1)})$, calculate the optimal decoding matrices ${\bf{U}}^{ (1)}$ and  auxiliary matrices ${\bf{W}}^{(1)}$ based on (\ref{fgrtgtyh});
\STATE Given ${\bf{U}}^{ (n)}$, ${\bf{W}}^{(n)}$ and ${\bm{\phi}}^{(n)}$, calculate the optimal  precoding matrices ${\bf{F}}^{(n+1)}$ by solving  Problem (\ref{appssxsorig}) using Algorithm \ref{itedasca};
 \STATE Given ${\bf{U}}^{ (n)}$, ${\bf{W}}^{(n)}$ and ${\bf{F}}^{(n+1)}$, calculate the optimal ${\bm{\phi}}^{(n+1)}$ by solving Problem (\ref{appsdsworig})  using Algorithm \ref{csdcaada};
 \STATE Given ${\bf{F}}^{(n+1)}$ and ${\bm{\phi}}^{(n+1)}$, calculate the optimal decoding matrices ${\bf{U}}^{ (n+1)}$ in (\ref{fgrtgtyh});
\STATE Given ${\bf{F}}^{(n+1)}$, ${\bf{U}}^{ (n+1)}$ and ${\bm{\phi}}^{(n+1)}$, calculate the optimal auxiliary matrices ${\bf{W}}^{(n+1)}$ in (\ref{fgrtgtyh});
 \STATE If $n\geq n_{\rm{max}}$ or ${{\left| {R({\bf{F}}^{(n+1)},{\bm{\phi}}^{(n+1)}) - R({\bf{F}}^{(n)},{\bm{\phi}}^{(n)})} \right|} \mathord{\left/
 {\vphantom {{\left| {{\rm{Obj}}({{\bf{F}}^{(n + 1)}},{{\bm{\theta}}^{(n + 1)}}) - R({\bf{F}}^{(n)},{\bm{\phi}}^{(n)})} \right|} {R({\bf{F}}^{(n)},{\bm{\phi}}^{(n)})}}} \right.
 \kern-\nulldelimiterspace} {R({\bf{F}}^{(n+1)},{\bm{\phi}}^{(n+1)})}} < \varepsilon$, terminate.  Otherwise, set $n \leftarrow n + 1$  and go to step 2.
\end{algorithmic}
\end{algorithm}

The following theorem shows the convergence and solution properties of Algorithm \ref{bcd}.

\itshape \textbf{Theorem 4:}  \upshape The OF value sequence $\{R({{\bf{F}}^{(n)}},{\bm{\phi}}^{(n)}{\rm{)}},n=1,2,\cdots\}$ generated by  Algorithm \ref{bcd} is guaranteed
to converge, and the final solution satisfies the KKT conditions of Problem (\ref{appstaoneorig}).

\itshape \textbf{Proof:}  \upshape Please refer to Appendix \ref{prooftheorem4}.  \hfill $\Box$

 The complexity of Algorithm \ref{bcd} mainly depends on that of Step 2 and Step 3, the complexity of which has been analyzed in the above subsections. In specific, the total complexity of step 2 and step 3 are respectively given by ${\cal O}(n_1^{\rm{max}}{\log _2}\left( {\frac{{{\lambda _u} - {\lambda _l}}}{\varepsilon }} \right) K_IN_B^3)$ and ${\cal O}(n_2^{\rm{max}}{\log _2}\left( {\frac{{{p _u} - {p _l}}}{\varepsilon }} \right) M^3)$, where $n_1^{\rm{max}}$ and $n_2^{\rm{max}}$ denote  the number of iterations for Algorithm \ref{itedasca} and Algorithm \ref{csdcaada} to converge. Denote the total number of iterations of Algorithm \ref{bcd} as $N_{\rm{max}}$.  Then, the overall complexity of Algorithm \ref{bcd} is given by $O\left( {{N_{\max }}\left( {n_1^{{\rm{max}}}{{\log }_2}\left( {\frac{{{\lambda _u} - {\lambda _l}}}{\varepsilon }} \right){K_I}N_B^3 + n_2^{{\rm{max}}}{{\log }_2}\left( {\frac{{{p_u} - {p_l}}}{\varepsilon }} \right){M^3}} \right)} \right)$. Additionally, the simulation results show that Algorithm  \ref{bcd} converges rapidly, which demonstrates the low complexity of this algorithm.

\section{Feasibility Check for Problem (\ref{appstaoneorig})}\label{frjgtoj}

Due to the conflicting EH  and limited transmit power constraints, Problem (\ref{appstaoneorig}) may be infeasible. Hence, we have to first check whether Problem (\ref{appstaoneorig}) is feasible or not. To this end, we construct the following optimization problem:
\begin{subequations}\label{aneorcdsig}
\begin{align}
&\mathop {\max }\limits_{{{\bf{F}},{\bm{\Phi}}  } } \
   {\rm{tr}}\left( {\sum\limits_{k = 1}^{K_I} {{\bf{F}}_k^{\rm{H}}{\bf{G}}{{\bf{F}}_k}} } \right)
\\
&\textrm{s.t.}\quad (\ref{dehwifr}), (\ref{jiofjj}).
\end{align}
\end{subequations}
If the optimal OF value is larger than $\bar Q$, Problem (\ref{appstaoneorig}) is feasible. Otherwise, it is infeasible. As the TPC matrices and phase shift matrix are coupled, the globally optimal solution is difficult to obtain. In the following, we can obtain a suboptimal solution by alternatively optimizing the TPC matrices and phase shifts.

For a given phase shift matrix, the TPC matrix optimization problem is given by
 \begin{subequations}\label{aneoasxsdsig}
\begin{align}
&\mathop {\max }\limits_{{{\bf{F}}} } \
   {\rm{tr}}\left( {\sum\limits_{k = 1}^{K_I} {{\bf{F}}_k^{\rm{H}}{\bf{G}}{{\bf{F}}_k}} } \right)
\\
&\textrm{s.t.}\quad (\ref{dehwifr}).
\end{align}
\end{subequations}
Upon denoting the maximum eigenvalue and the corresponding eigenvector of ${\bf{G}}$ by $\chi$ and $\bf{b}$ respectively, the optimal solution can be readily obtained as ${{\bf{F}}_k} = \left[ {\sqrt {{p_k}} {\bf{b}},{{\bf{0}}_{{N_B} \times (d - 1)}}} \right],\forall k=1,\cdots,K_I$, where $\sum\nolimits_{k = 1}^{{K_I}} {{p_k}}  = {P_T}$ and $p_k\ge 0, \forall k=1,\cdots,K_I$. Without loss of generality, we can set  $p_i=P_T/K_I,\forall i\in {\cal K}_I$. The OF value is given by $\chi P_T$. In this case, the optimal TPC matrix represents the optimal energy beamforming, which is the same as that for the single-antenna IR case of \cite{xu2014multiuser}.

For a given TPC matrix $\bf{F}$, the phase shift optimization problem is formulated as:
 \begin{subequations}\label{anasxssig}
\begin{align}
&\mathop {\max }\limits_{{{\bm{\phi}}} } \
   {{\bm{\phi}} ^{\rm{H}}}\bm{\Upsilon}{\bm{\phi}}+2{\mathop{\rm Re}\nolimits} \left\{ {{{\bm{\phi}} ^{\rm{H}}}{\bf{g}^*}} \right\}
\\
&\textrm{s.t.}\quad (\ref{jiofjj}),
\end{align}
\end{subequations}
where $\bm{\Upsilon}$ and $\bf{g}$ are defined in the above section.
The OF is   convex  w.r.t. ${\bm{\phi}}$, and maximizing a convex function is a d.c program. Hence, it can be solved by using the SCA method by approximating $ {{\bm{\phi}} ^{\rm{H}}}\bm{\Upsilon}{\bm{\phi}}$ as its first-order Taylor expansion, details of which are omitted.

Finally, alternatively solve Problem (\ref{aneoasxsdsig}) and (\ref{anasxssig}) until the OF is larger than $\bar Q$.

\section{Simulation Results}\label{simlresult}
\begin{figure}
\centering
\includegraphics[width=3.4in]{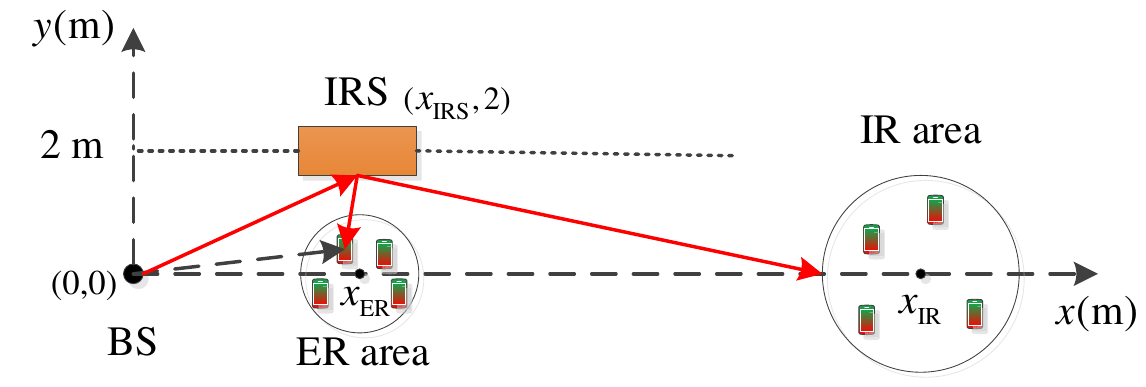}\vspace{-0.2cm}
\caption{The simulated  IRS-aided SWIPT MIMO communication scenario. }
\label{fig2}\vspace{-0.5cm}
\end{figure}

In this section, we provide simulation results for demonstrating the benefits of applying IRS to SWIPT  systems, as seen in Fig.~\ref{fig2}, where there are four ERs and two IRs. The ERs and IRs are uniformly and randomly scattered in a circle centered at $(x_{\rm{ER}},0)$ and $(x_{\rm{IR}},0)$ with radius 1 m and 4 m, respectively. The IRS is located at $(x_{\rm{IRS}}, 2)$. In the simulations, we assume that the IRS is just above the ERs and thus we set $x_{\rm{ER}}=x_{\rm{IRS}}$. The large-scale path loss is modeled in dB as
\begin{equation}\label{scfadWDE}
 PL = P{L_0}{\left( {\frac{D}{{{D_0}}}} \right)^{ - \alpha }},
\end{equation}
where ${\rm{P}}{{\rm{L}}_0}$ is the path loss at the reference distance $D_0$,   $D$ is the link length in meters, and $\alpha$ is the path loss exponent. Here, we set $D_0=1$ and ${\rm{P}}{{\rm{L}}_0}=-30 \rm{dB}$.  The path loss exponents of the BS-IRS, IRS-ER, IRS-IR, BS-IR and BS-ER links are respectively set as $\alpha_{\rm{BSIRS}}=2.2$, $\alpha_{\rm{IRSER}}=2.2$, $\alpha_{\rm{IRSIR}}=2.4$, $\alpha_{\rm{BSIR}}=3.6$ and $\alpha_{\rm{BSER}}=3.6$. Unless otherwise stated, the other parameters are set as follows:
 Channel bandwidth of 1 MHz, noise power density of $-160$ dBm/Hz, $N_B=4$, $N_I=N_E=2$, $d=2$, $\bar Q=2 \times {10^{ - 4}}$ W, $\eta=0.5$, $M=50$, $P_T=10\  W$, weight factors $\omega_k=1, \forall k\in {\cal K}_I, {\alpha _l}=1, \forall l\in {\cal K}_E$, $x_{\rm{ER}}=5 $ m, and $x_{\rm{IR}}=400 $ m. The following results are obtained by averaging over 100 random locations and channel generations. Due to the severe blockage and long distance, the channels from the BS and the IRS to the IRs are assumed to be Rayleigh fading. However, as the BS, the ERs and the IRS are close to each other, the small-scale channels are assumed to be  Rician fading. In particular, the small-scale channels from the IRS to the ERs are denoted as:
 \begin{equation}\label{fvgbhnjik}
  {{\bf{\tilde G}}_{r,l}} = \sqrt {\frac{{{\beta _{{\rm{irser}}}}}}{{{\beta _{{\rm{irser}}}} + 1}}} {\bf{\tilde G}}_{r,l}^{{\rm{LoS}}} + \sqrt {\frac{1}{{{\beta _{{\rm{irser}}}} + 1}}} {\bf{\tilde G}}_{r,l}^{{\rm{NLoS}}},l = 1, \cdots ,{K_E},
 \end{equation}
 where  $\beta _{{\rm{irser}}}$ is the Rician factor, ${\bf{\tilde G}}_{r,l}^{{\rm{LoS}}}$ is the deterministic line of sight (LoS), and ${\bf{\tilde G}}_{r,l}^{{\rm{NLoS}}}$ is the non-LoS (NLoS) component that is  Rayleigh fading. The LoS component
 ${\bf{\tilde G}}_{r,l}^{{\rm{LoS}}}$ can be modeled as  ${\bf{\tilde G}}_{r,l}^{{\rm{LoS}}} = {{\bf{a}}_{{N_E}}}\left( {\vartheta _{{\rm{irser}},l}^{AoA}} \right){\bf{a}}_M^H\left( {\vartheta _{{\rm{irser}},l}^{AoD}} \right)$, where ${{\bf{a}}_{{N_E}}}\left( {\vartheta _{{\rm{irser}},l}^{AoA}} \right)$  is defined as
 \begin{equation}\label{tgthbty}
  {{\bf{a}}_{{N_E}}}\left( {\vartheta _{{\rm{irser}},l}^{AoA}} \right) = {\left[ {1,{e^{j\frac{{2\pi d}}{\lambda }\sin \vartheta _{{\rm{irser}},l}^{AoA}}}, \cdots ,{e^{j\frac{{2\pi d}}{\lambda }({N_E} - 1)\sin \vartheta _{{\rm{irser}},l}^{AoA}}}} \right]^T}
 \end{equation}
 and
\begin{equation}\label{jpjookko}
  {{\bf{a}}_M}\left( {\vartheta _{{\rm{irser}},l}^{AoD}} \right) = {\left[ {1,{e^{j\frac{{2\pi d}}{\lambda }\sin \vartheta _{{\rm{irser}},l}^{AoD}}}, \cdots ,{e^{j\frac{{2\pi d}}{\lambda }(M - 1)\sin \vartheta _{{\rm{irser}},l}^{AoD}}}} \right]^T}.
\end{equation}
In (\ref{tgthbty}) and (\ref{jpjookko}), $d$ is the antenna separation distance, $\lambda$ is the wavelength, $\vartheta _{{\rm{irser}},l}^{AoD}$ is the angle of departure and $\vartheta _{{\rm{irser}},l}^{AoA}$ is the angle of arrival. It is assumed that $\vartheta _{{\rm{irser}},l}^{AoD}$ and  $\vartheta _{{\rm{irser}},l}^{AoA}$ are randomly distributed within $[0,2\pi ]$. For simplicity, we set $d/\lambda=1/2$. The small-scale channels from the BS to the ERs and the IRS are similarly defined. For simplicity, the Rician factors for all Rician fading channels are assumed to be the same as $\beta=3$.

\begin{figure}
\begin{minipage}[t]{0.495\linewidth}
\centering
\includegraphics[width=2.6in]{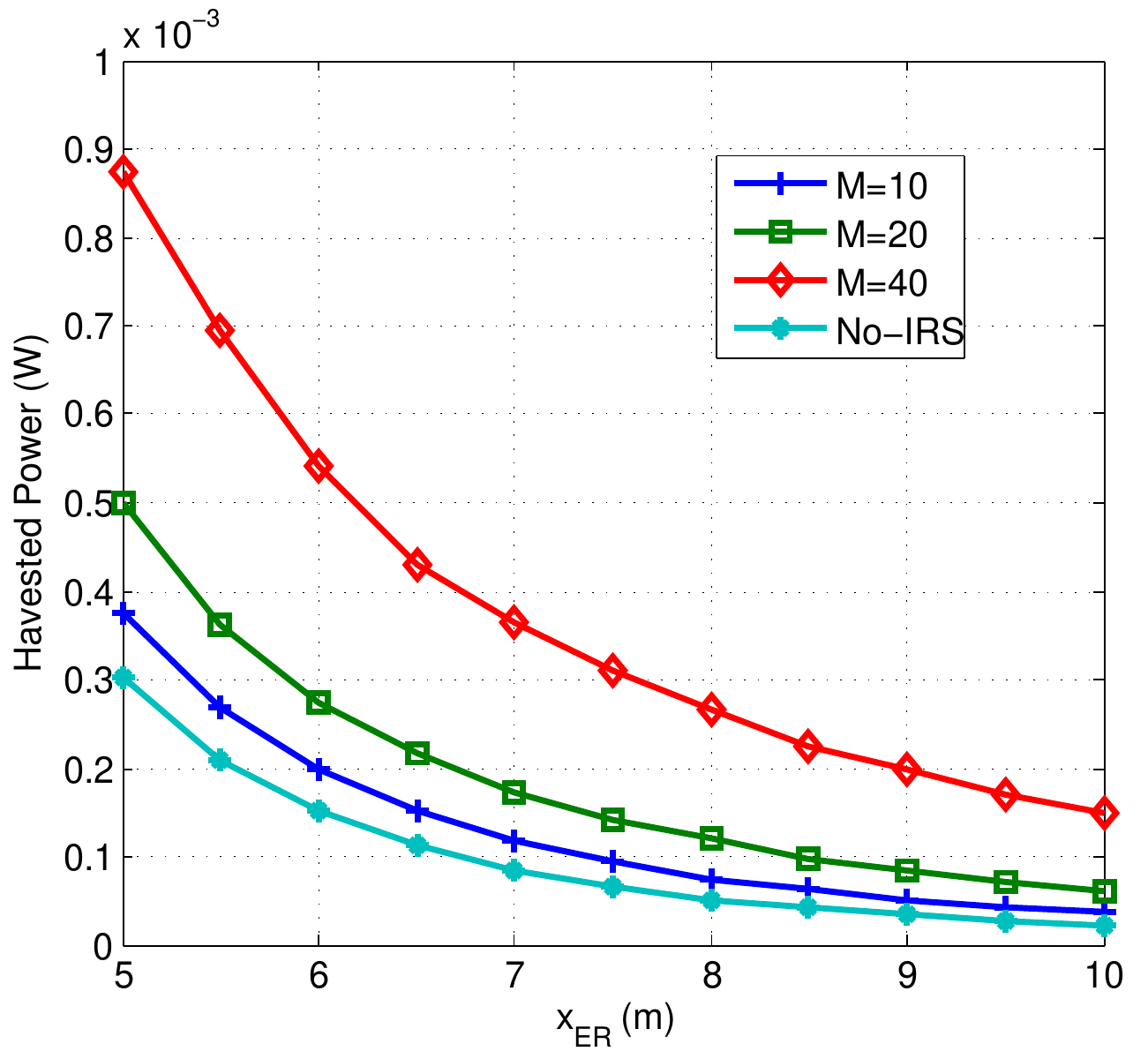}\vspace{-0.6cm}
\caption{Maximum harvested power achieved by various schemes.}
\label{fig3}
\end{minipage}%
\hfill
\begin{minipage}[t]{0.495\linewidth}
\centering
\includegraphics[width=2.6in]{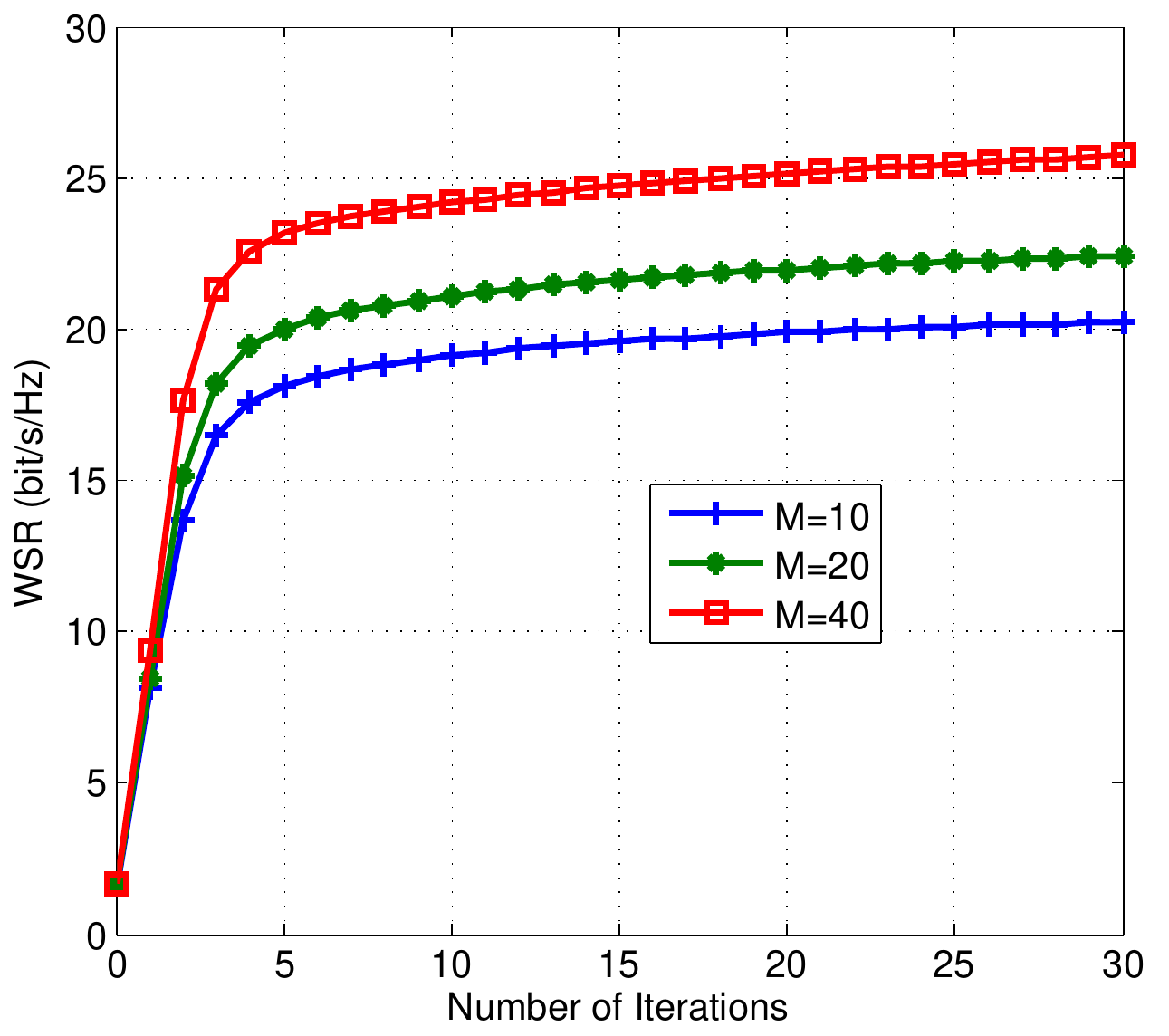}\vspace{-0.6cm}
\caption{Convergence behaviour of the BCD algorithm.}
\label{fig4}
\end{minipage}\vspace{-0.7cm}
\end{figure}

We first compare the maximum  power harvested   by various schemes in Fig.~\ref{fig3}. Specifically, we solve the EH maximization problem (\ref{aneorcdsig}) by using the feasibility check method in Section \ref{frjgtoj}. Additionally, we also present the results without using IRS. Fig.~\ref{fig3} shows the maximum EH power versus the ER circle center location $x_{\rm{EH}}$. As expected, the EH power gleaned by all schemes decreases, when the ERs are far away from the BS. As expected, more power can be harvested with the aid of IRS than that without IRS, especially when the number of phase shifters $M$ is large. This is mainly due to the fact that an additional strong link is reflected by the IRS, which can be harvested by the ERs. This figure also shows that the IRS is effective in expanding the operational range of ERs. For example, when the harvested power limit is $\bar Q=2\times 10^{-4}$ W, the maximum operational range of the system without IRS is only 5.5 m, while the system having $M=40$ phase shifters can operate for distances up to 9 m.

In Fig.~\ref{fig4}, we study the convergence behaviour of the BCD algorithm for different numbers of phase shifters $M$. It is observed from Fig.~\ref{fig4} that the WSR achieved for various $M$ values increases monotonically  with the number of iterations, which verifies Theorem 4. Additionally, the BCD algorithm converges rapidly and in general a few iterations are sufficient for the BCD algorithm to achieve a large portion of the converged WSR. This reflects the low complexity of the BCD algorithm, which is appealing for practical applications.

\begin{figure}
\begin{minipage}[t]{0.495\linewidth}
\centering
\includegraphics[width=2.6in]{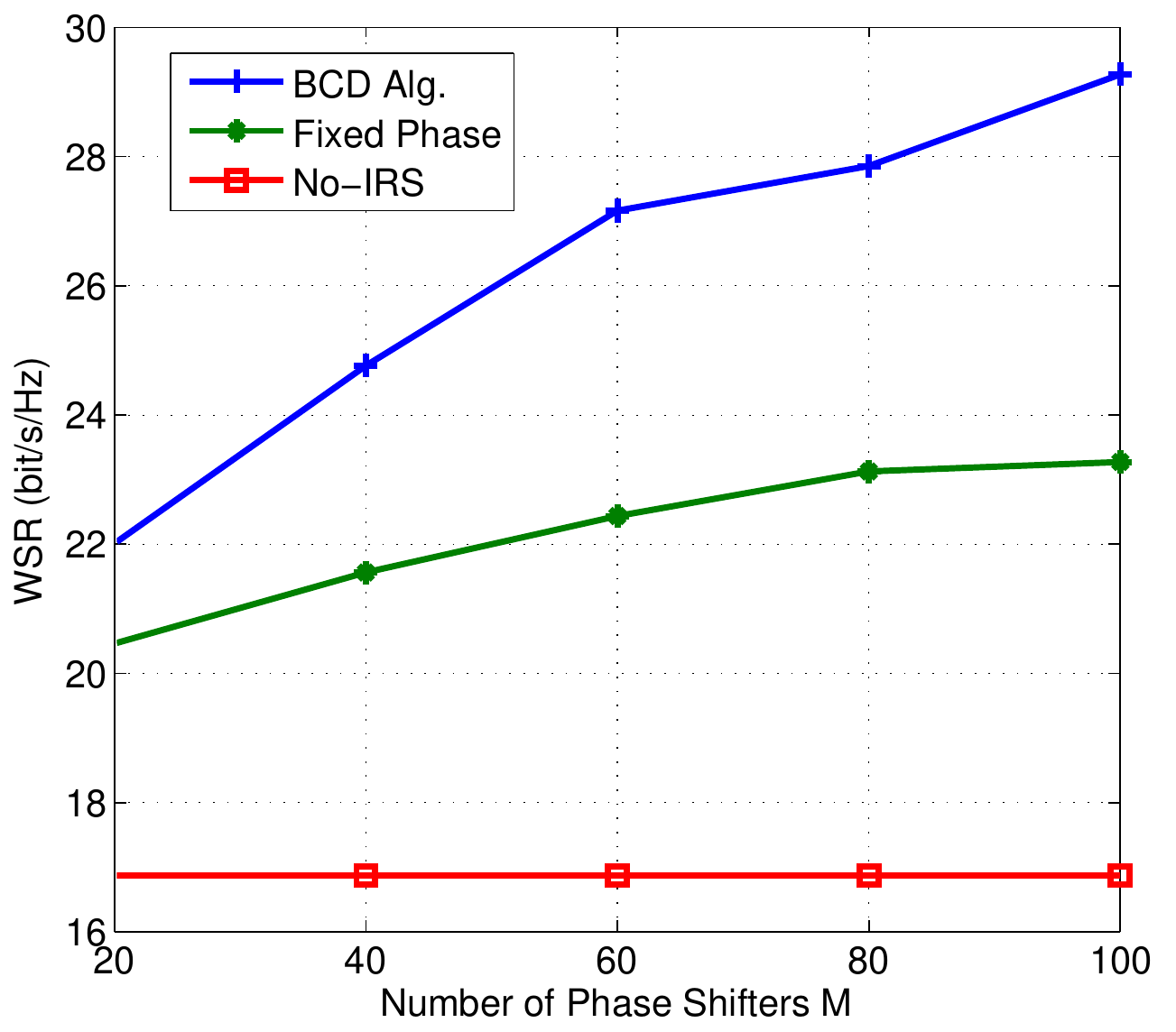}\vspace{-0.6cm}
\caption{ WSR versus the number of phase shifters.}
\label{fig5}
\end{minipage}%
\hfill
\begin{minipage}[t]{0.495\linewidth}
\centering
\includegraphics[width=2.6in]{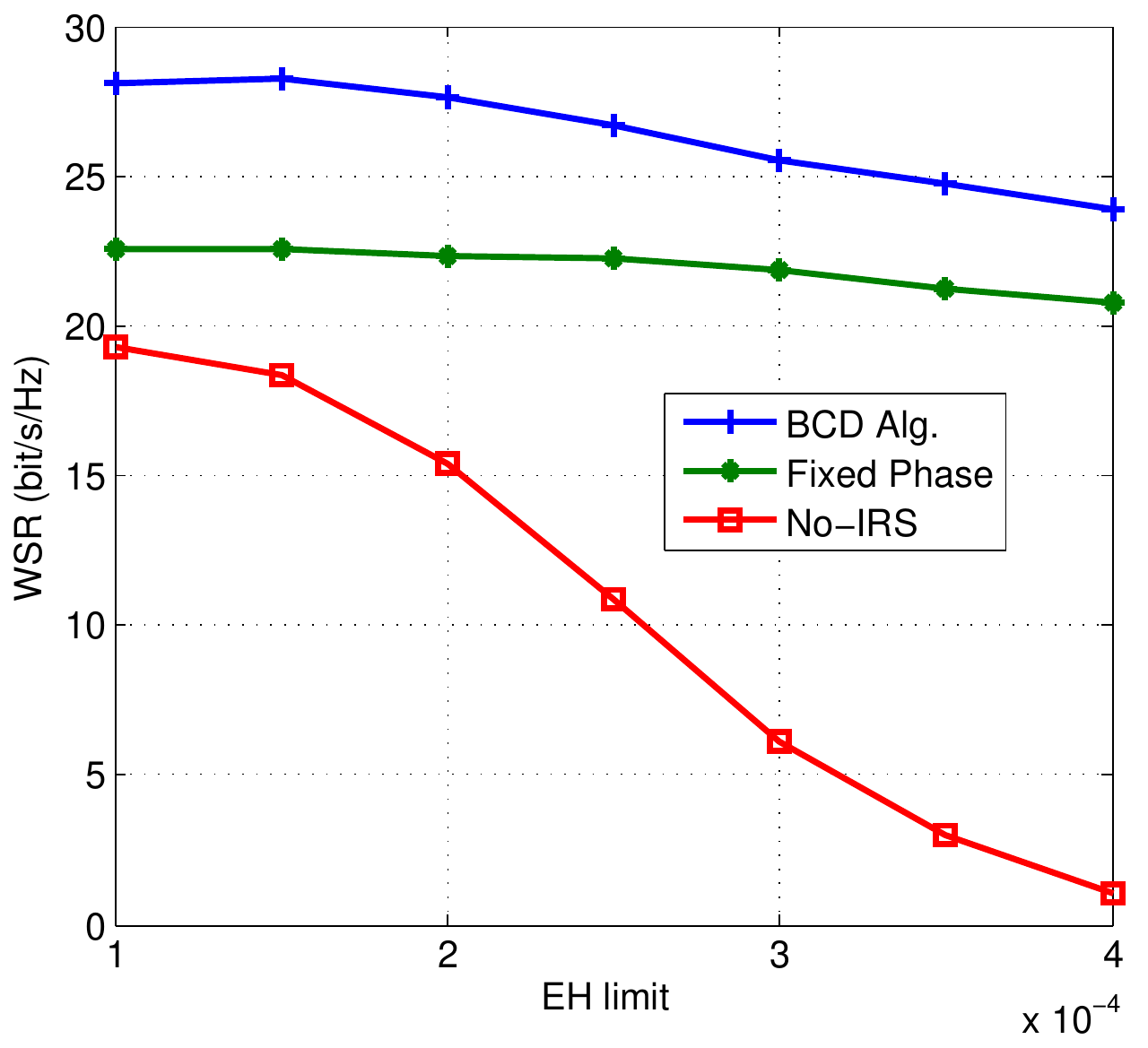}\vspace{-0.6cm}
\caption{WSR versus the harvested power requirement $\bar Q$.}
\label{fig6}
\end{minipage}\vspace{-0.7cm}
\end{figure}

In the following, we compare our proposed BCD algorithms to a pair of benchmark schemes: 1)`No-IRS': In this scheme, there is no IRS to assist the transmission as in conventional systems; 2) `Fixed Phase': In this method, the phase shifts are fixed at the solutions obtained by solving the harvested power maximization problem (\ref{aneorcdsig}), while they are not optimized, when using the BCD algorithm by removing Step 3 of the BCD algorithm. When any of the methods fails to obtain a feasible solution, its achievable WSR is set to zero.

In Fig.~\ref{fig5}, we first study the impact of the number of phase shifters $M$ on the performance of various algorithms. As expected, the WSR achieved by all the algorithms - except for the No-IRS method - increases with $M$, since a higher degree of freedom can be exploited for optimizing the system performance. By carefully optimizing the phase shifts for maximizing the WSR, the BCD algorithm significantly outperforms   the fixed-phase scheme. Additionally, the performance gain increases with $M$, which emphasizes the importance of optimizing the phase shifts. By employing the IRS in our SWIPT system, the WSR obtained by the BCD algorithm becomes drastically higher than that of No-IRS. For example, when $M=60$, the WSR performance gain is up to 10 bit/s/Hz. These results demonstrate that introducing the IRS into our SWIPT system is a   promising technique of  enhancing the system performance.

In Fig.~\ref{fig6}, the impact of harvested power requirement $\bar Q$ is investigated. It is seen from this figure that the WSR achieved by all the algorithms decreases upon increasing $\bar Q$,  because the probability of infeasibility increases, which in turn  reduces the average WSR value. We also find that the WSR obtained by the No-IRS scheme decreases more rapidly than that of the other two IRS-aided transmission schemes. The WSR of the No-IRS is approaching zero when $\bar Q=4\times 10^{-4}$ W, while those relying on IRSs   achieve  a WSR gain in excess of 20 bit/s/Hz. It is  observed again that the BCD algorithm performs better than the fixed-phase scheme, but the gap narrows with the increase of $\bar Q$. This can be explained as follows. With the increase of $\bar Q$,  both the TPC matrices and the phase shifts should be designed for maximizing the  power harvested at the ERs, and thus the final solutions of the fixed-phase and BCD method will become the same.

\begin{figure}
\begin{minipage}[t]{0.495\linewidth}
\centering
\includegraphics[width=2.6in]{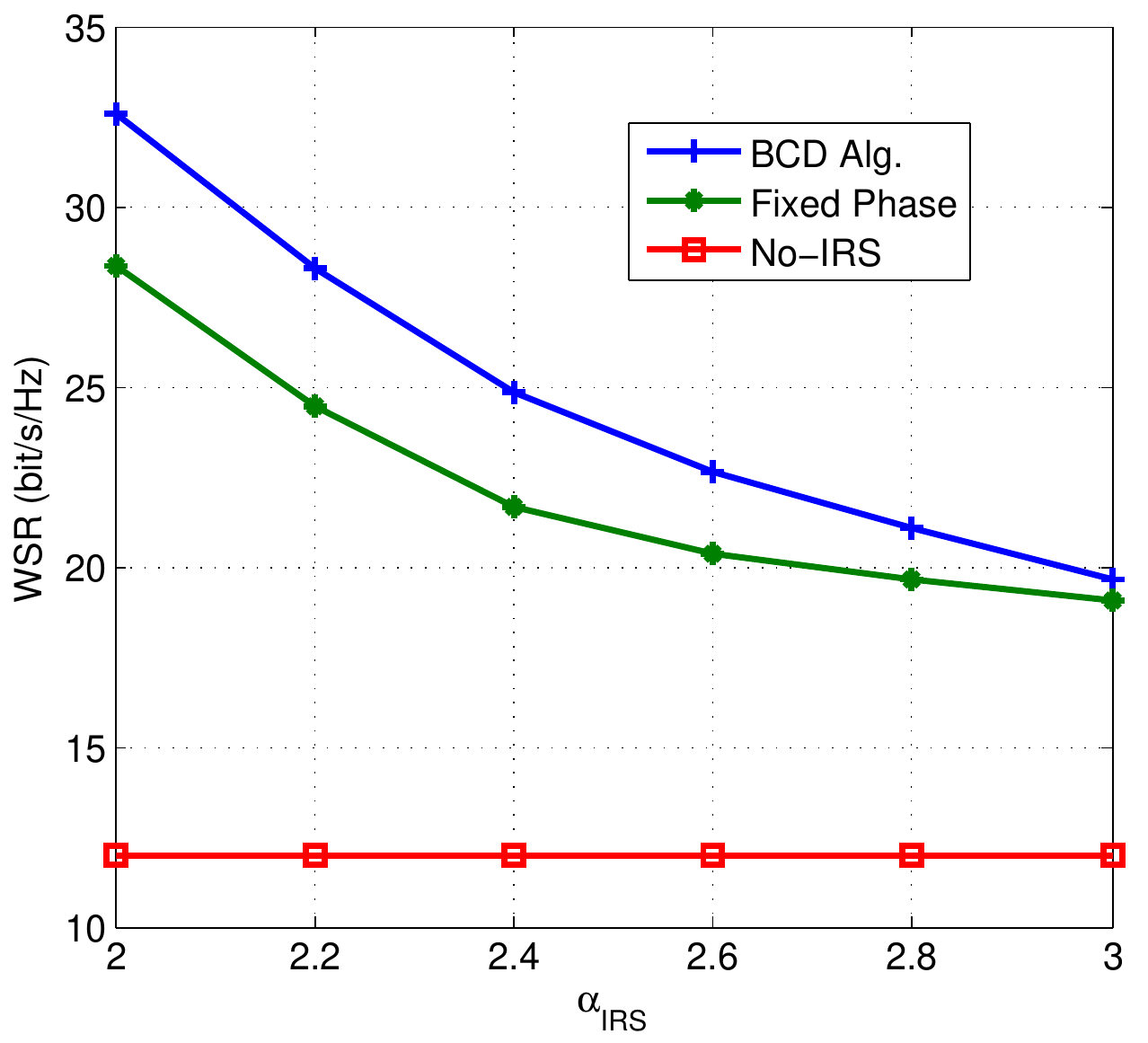}\vspace{-0.6cm}
\caption{ WSR versus the IRS-related path loss exponent $\alpha_{\rm{IRS}}$.}
\label{fig7}
\end{minipage}%
\hfill
\begin{minipage}[t]{0.495\linewidth}
\centering
\includegraphics[width=2.6in]{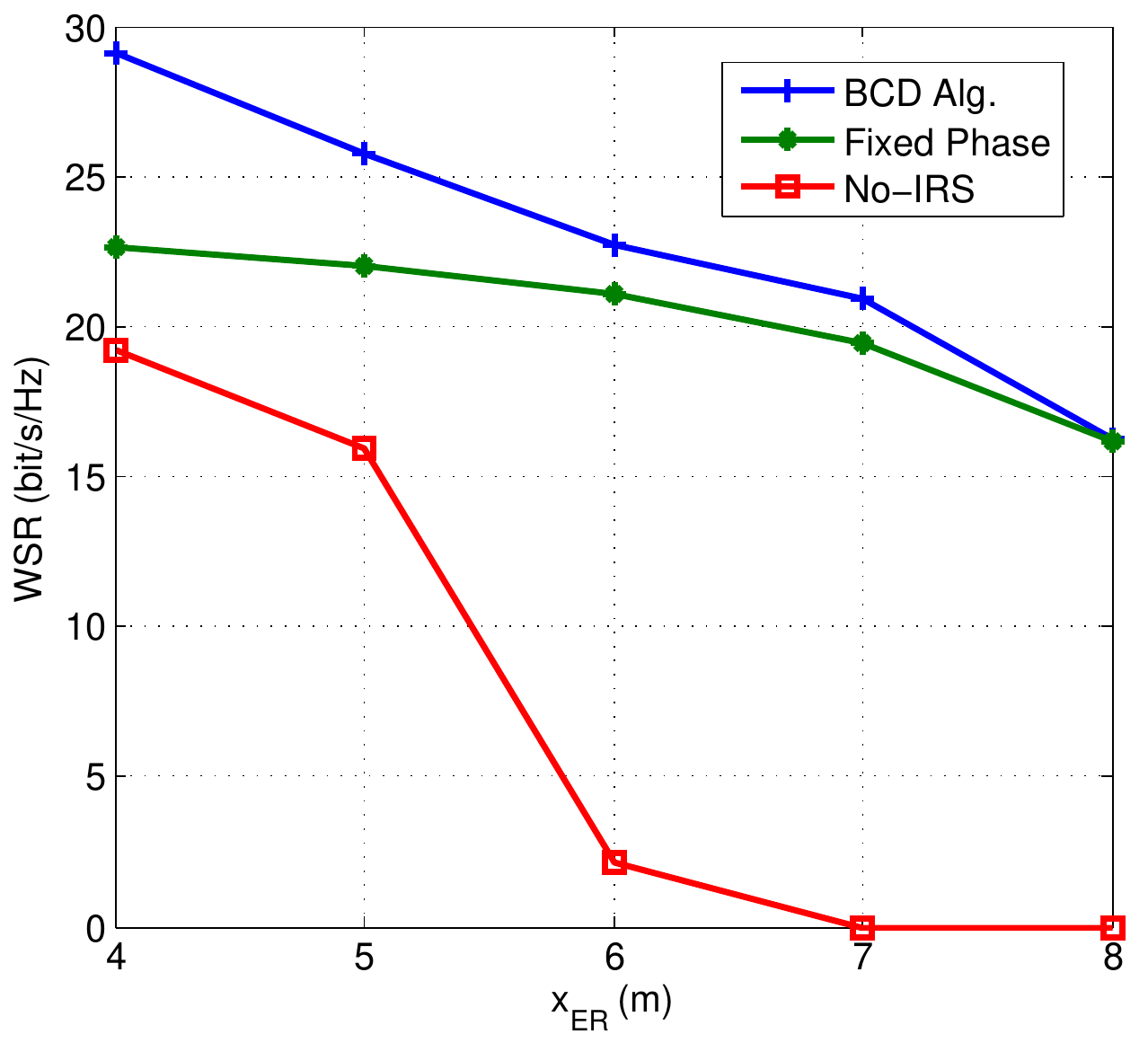}\vspace{-0.6cm}
\caption{WSR versus the location of ER circle center $x_{\rm{ER}}$.}
\label{fig8}
\end{minipage}\vspace{-0.7cm}
\end{figure}

The above results are obtained for $\alpha_{\rm{BSIRS}}=2.2$, $\alpha_{\rm{IRSER}}=2.2$, $\alpha_{\rm{IRSIR}}=2.4$ based on the assumption that the IRS relies on an obstacle-free scenario. In practice, this ideal scenario is seldom encountered. Hence, it is imperative to investigate the impact of $\alpha_{\rm{IRS}}\buildrel \Delta \over =\alpha_{\rm{BSIRS}}=\alpha_{\rm{IRSER}}=\alpha_{\rm{IRSIR}}$ on the system performance, which is shown in Fig.~\ref{fig7}. Observe from this figure that the WSR achieved by the algorithms using IRS decreases drastically with $\alpha_{\rm{IRS}}$. When $\alpha_{\rm{IRS}}=3$, the WSR-performance gain of our algorithm over the No-IRS scenario is only 7 bit/s/Hz, because upon increasing  $\alpha_{\rm{IRS}}$, the signal power reflected from the IRS becomes weaker. Hence, the benefits of  the IRS can be eroded. This provides an important engineering design insight: the location of IRS should be carefully considered for finding an obstacle-free scenario associated with a low $\alpha_{\rm{IRS}}$.

In Fig.~\ref{fig8}, we study the impact of ER locations on the system performance. As expected, the WSR achieved by all the schemes decreases with $x_{\rm{IRS}}$, since the ERs become more distant  from the BS and the signals gleaned from both the BS and IRS become weaker. The WSR achieved by the No-IRS approaches zero when $x_{\rm{IRS}}=8$ m, hence this method cannot reach the energy transmission target of the ERs. The proposed algorithm is again observed to significantly outperform the other two algorithms, especially when the ERs are close to the BS.
\begin{figure}
  \centering
  \includegraphics[width=3in]{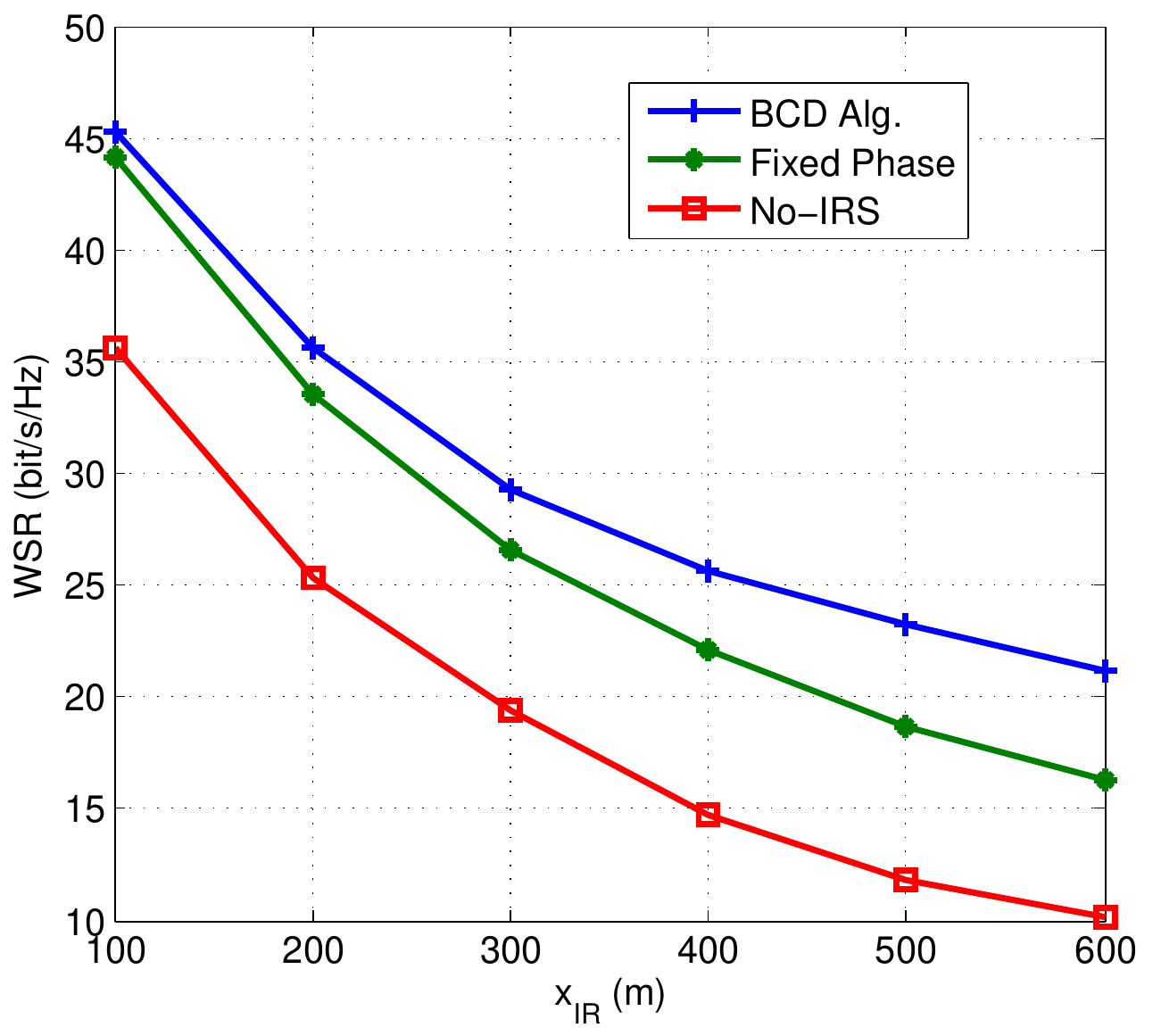}\\
  \caption{ WSR versus the location of IR circle center $x_{\rm{IR}}$. }\label{fig9}
\end{figure}

 Finally, the impact of IR locations is investigated in Fig.~\ref{fig9}. It is observed that the WSR achieved by all the algorithms decreases with $x_{\rm{IR}}$ since the IRs become farther away from the BS when increasing $x_{\rm{IR}}$. The proposed algorithm is shown to achieve nearly the WSR gain of 10 bit/s/Hz over the No-IRS when $x_{\rm{IR}}=100\ m$, and the WSR gain  slightly increases with $x_{\rm{IR}}$. This means that the IRS is more advantageous when the IRs are far away from the BS, and the IRS can provide one additional favorable link.

\section{Conclusions}\label{conclu}

In this paper, we have invoked  an IRS in a SWIPT MIMO system for enhancing   the performance of both the ERs and IRs. By carefully adjusting the phase shifts at the IRS, the  signal reflected by the IRS can be added constructively at both the ERs and IRs. We considered the WSR maximization problem of IRs, while guaranteeing the energy harvesting requirements of the ERs and the associated non-convex unit-modulus constraints. We conceived a BCD algorithm for alternatively optimizing the TPC matrices at the BS and the phase shift matrix at the IRSs. For each subproblem, a low-complexity iterative algorithm was proposed, which   guarantees to be at worst locally optimal. Our simulation results demonstrated that the IRS  enhances  the performance of the SWIPT system  and that the proposed algorithm converges rapidly, hence it is eminently suitable for practical implementations.

 This paper assumes perfect CSI at the BS, which is challenging to obtain. For the future work, we will consider the robust transmission design for the IRS-aided SWIPT system, where the CSI is assumed to be imperfectly known. In addition, how to design the discrete phase shifts will be left for future work.
\numberwithin{equation}{section}
\begin{appendices}

\section{Proof of Lemma 1}\label{prooflemma1}

We consider a pair of variables $\lambda$ and $\lambda'$, where $\lambda>\lambda'$. Let  ${\bf{F}(\lambda )}$ and  ${\bf{F}(\lambda')}$ be the optimal solutions of Problem (\ref{adwef}) with $\lambda$ and $\lambda'$, respectively. Since ${\bf{F}(\lambda )}$ is the optimal solution of Problem (\ref{adwef}) with $\lambda $, we have
\vspace{-0.2cm}
\begin{equation}\label{dafrer}
  {\cal L}[{\bf{F}(\lambda )},\lambda ] \le {\cal L}[{\bf{F}(\lambda' )},\lambda ].
\end{equation}
\vspace{-0.2cm}
Similarly, we have
\begin{equation}\label{dasdr}
  {\cal L}[{\bf{F}(\lambda' )},\lambda' ]\le {\cal L}[{\bf{F}(\lambda )},\lambda' ].
\end{equation}
By adding these two inequalities and simplifying them, we have $\left( {\lambda  - \lambda '} \right)P(\lambda ) \le \left( {\lambda  - \lambda '} \right)P(\lambda ')$. Since $\lambda>\lambda'$, we have $P(\lambda ) \le P(\lambda ')$, which completes the proof.

\vspace{-0.3cm}\section{Proof of Theorem 2}\label{prooftheorem2}

Denote the globally optimal solution of Problem (\ref{asasdcjig}) by $\bm{\phi}^\star$. According to \cite{boyd2004convex}, for a non-convex optimization problem, all its locally optimal solutions (including the globally optimal solution) should satisfy the Karush-Kuhn-Tucker (KKT) optimality conditions, one of which is the complementary slackness condition for constraint (\ref{sacvfgatr}):
\vspace{-0.2cm}
 \begin{equation}\label{axdewfw}
 \lambda^\star \left(2{\mathop{\rm Re}\nolimits} \left[ {{{\bm{\phi}} ^{\star\rm{H}}}\left( {{\bf{g}^*} + {\bm{\Upsilon} }{{\bm{\phi}} ^{(n) }}} \right)} \right] -\hat Q\right)=0,  \vspace{-0.2cm}
 \end{equation}
 where $ \lambda^\star$ is the corresponding optimal dual variable. We consider two cases: 1) $ \lambda^\star=0$; 2) $ \lambda^\star>0$.

The first case means that constraint (\ref{sacvfgatr}) is not tight in the optimum. Then, the optimal solution can be obtained as $\bm{\phi}^\star={e^{j\arg \left({\bf{q}}^{(n)}\right)}}$, which is equal to $\bm{\phi}{(0)}$. Hence, Algorithm \ref{isaaada}
 achieves the optimal solution of Problem  (\ref{asasdcjig}).

For the second case, the following equality should hold:
\begin{equation}\label{wdferf}
   2{\mathop{\rm Re}\nolimits} \left[ {{{\bm{\phi}} ^{\star\rm{H}}}\left( {{\bf{g}^*} + {\bm{\Upsilon} }{{\bm{\phi}} ^{(n) }}} \right)} \right]=\hat Q.
\end{equation}
We prove the second case by using the  method of contradiction. Denote the optimal $p$ obtained by Algorithm \ref{isaaada} as $p^\star$, and the corresponding $\bm{\phi}$ as $\bm{\phi}(p^\star)$. Then, we have
\begin{equation}\label{wsqdferfr}
  2{\mathop{\rm Re}\nolimits} \left[ {{\bm{\phi}(p^\star) ^{\rm{H}}}\left( {{\bf{g}^*} + {\bm{\Upsilon} }{{\bm{\phi}} ^{(n) }}} \right)} \right]=\hat Q.
\end{equation}
Let us assume that $\bm{\phi}(p^\star)$ is not the globally optimal solution of Problem  (\ref{asasdcjig}). Then, we have
\begin{equation}\label{aceffr}
 2{\mathop{\rm Re}\nolimits} \left\{ {{\bm{\phi}(p^\star) ^{\rm{H}}{\bf{q}}^{(n)}} } \right\}<2{\mathop{\rm Re}\nolimits} \left\{ {{\bm{\phi}^{\star\rm{H}}{\bf{q}}^{(n)}} } \right\}.
\end{equation}
Since $\bm{\phi}(p^\star)$ is the globally optimal solution of Problem (\ref{asasxwdwcjig}) when $p=p^\star$, we have
 \vspace{-0.2cm}
\begin{equation}\label{aerfrceffr}
\!\!\!  2{\mathop{\rm Re}\nolimits} \left\{ {{{\bm{\phi}}(p^\star) ^{\rm{H}}{\bf{q}}^{(n)}} } \right\}\!+\!2p^\star{\mathop{\rm Re}\nolimits} \left[ {{{\bm{\phi}}(p^\star) ^{\rm{H}}}\left( {{\bf{g}^*} \!+\! {\bm{\Upsilon} }{{\bm{\phi}} ^{(n) }}} \right)} \right]\!\ge\!2{\mathop{\rm Re}\nolimits} \left\{ {{{\bm{\phi}} ^{\star\rm{H}}{\bf{q}}^{(n)}} } \right\}\!+\!2p^\star{\mathop{\rm Re}\nolimits} \left[ {{{\bm{\phi}} ^{\star\rm{H}}}\left( {{\bf{g}^*} \!+\! {\bm{\Upsilon} }{{\bm{\phi}} ^{(n) }}} \right)} \right]. \vspace{-0.2cm}
\end{equation}
By substituting (\ref{wdferf}) and (\ref{wsqdferfr}) into (\ref{aerfrceffr}), we have
 \vspace{-0.2cm}
\begin{equation}\label{saxdef}
  2{\mathop{\rm Re}\nolimits} \left\{ {{{\bm{\phi}}(p^\star) ^{\rm{H}}{\bf{q}}^{(n)}} } \right\}\ge 2{\mathop{\rm Re}\nolimits} \left\{ {{{\bm{\phi}} ^{\star\rm{H}}{\bf{q}}^{(n)}} } \right\}, \vspace{-0.2cm}
\end{equation}
which contradicts (\ref{aceffr}). Hence, the solution obtained by Algorithm \ref{isaaada} is the globally optimal solution of Problem (\ref{asasdcjig}). Since Problem (\ref{apsasdcjig}) is equivalent to Problem (\ref{asasdcjig}), the proof is complete.

\vspace{-0.4cm}\section{Proof of Theorem 3}\label{prooftheorem3}
Let us define the following functions:
 \vspace{-0.2cm}
\begin{eqnarray}
  T({\bm{\phi}}) &\buildrel \Delta \over=&{{\bm{\phi}} ^{\rm{H}}}\bm{\Upsilon}{\bm{\phi}}+2{\mathop{\rm Re}\nolimits} \left\{ {{{\bm{\phi}} ^{\rm{H}}}{\bf{g}^*}} \right\} + {\rm{tr}}\left( {{{\bf{G}}_b}{\bf{\tilde F}}} \right),\\
 \bar T({\bm{\phi}}|{\bm{\phi}}^{(n)}) &\buildrel \Delta \over=& -{{\bm{\phi}^{(n) }} ^{\rm{H}}}\bm{\Upsilon}{\bm{\phi}^{(n) }}+ 2{\mathop{\rm Re}\nolimits} \left[ {{{\bm{\phi}} ^{\rm{H}}}\left( {{\bf{g}^*} + {\bm{\Upsilon} }{{\bm{\phi}} ^{(n) }}} \right)} \right]+ {\rm{tr}}\left( {{{\bf{G}}_b}{\bf{\tilde F}}} \right).
\end{eqnarray}
 \vspace{-0.2cm}
It can be verified that $T({\bm{\phi}}^{(n)})=\bar T({\bm{\phi}^{(n)}}|{\bm{\phi}}^{(n)})$.

We first show that the solution sequence $\{{\bm{\phi}}^{(n)},n=1,2,\cdots\}$ is feasible for Problem  (\ref{appsdsworig}). The unit-modulus constraint is guaranteed in (\ref{asxwed}). We only have to check the EH constraint in (\ref{hhudiuwe}). Note that ${\bm{\phi}}^{(n+1)}$ is a feasible solution of Problem  (\ref{asasdcjig}), and thus satisfies constraint (\ref{xaxasxaefe}). Hence, we have $\bar T({\bm{\phi}}^{(n+1)}|{\bm{\phi}}^{(n)})\ge \bar Q$. By using inequality (\ref{xaxasxaefe}), we have $T({\bm{\phi}}^{(n+1)})\ge \bar T({\bm{\phi}}^{(n+1)}|{\bm{\phi}}^{(n)})$. Then, $T({\bm{\phi}}^{(n+1)})\ge \bar Q$ holds, which means that the sequence of ${\bm{\phi}}^{(n+1)}$  satisfies the EH constraint in (\ref{hhudiuwe}).

Now, we show that the  OF value sequence $\{f({\bm{\phi}}^{(n)}),n=1,2,\cdots\}$ is monotonically decreasing. Based on Theorem 2, the globally optimal solution $\bm{\Phi}$ to Problem  (\ref{apsasdcjig})
 can be obtained. Then, we have $g({\bm{\phi}}^{(n+1)}|{\bm{\phi}}^{(n)})\le g({\bm{\phi}^{(n)}}|{\bm{\phi}}^{(n)})$. According to the first condition in (\ref{adefe}), we have  $g({\bm{\phi}}^{(n)}|{\bm{\phi}}^{(n)})\!=\!f({\bm{\phi}}^{(n)})$. Hence, we have $g({\bm{\phi}}^{(n+1)}|{\bm{\phi}}^{(n)})\le f({\bm{\phi}}^{(n)})$. By using the third condition of (\ref{adefe}), we have $g({\bm{\phi}}^{(n+1)}|{\bm{\phi}}^{(n)})\ge f({\bm{\phi}}^{(n+1)})$. As a result, we have $f({\bm{\phi}}^{(n)})\ge f({\bm{\phi}}^{(n+1)})$. Additionally, the OF must have a lower bound due to the unit-modulus constraint. Hence, the OF value sequence $\{f({\bm{\phi}}^{(n)}),n=1,2,\cdots\}$ is guaranteed to converge.

 Now, we prove that the converged solution satisfies the KKT conditions of Problem  (\ref{appsdsworig}). Let us denote the converged solution by $\{{\bm{\phi}}^\star \}$. Since ${\bm{\phi}}^\star$ is the globally optimal solution of Problem (\ref{apsasdcjig}), it must satisfy the KKT conditions of Problem (\ref{apsasdcjig}). Specifically, the Lagrange function of Problem (\ref{apsasdcjig}) is given by
  \vspace{-0.2cm}
\begin{equation}\label{safere}
{\cal L}({\bm{\phi}},\nu, \bm{\tau} )= g({\bm{\phi}}|{\bm{\phi}}^\star)+ \nu\left( \hat Q-2{\mathop{\rm Re}\nolimits} \left[ {{{\bm{\phi}} ^{\rm{H}}}\left( {{\bf{g}^*} + {\bm{\Upsilon} }{{\bm{\phi}} ^\star}} \right)} \right]\right)+\sum\limits_{m = 1}^M {{\tau _m}\left( {\left| {{\phi _m}} \right| - 1} \right)}, \vspace{-0.2cm}
\end{equation}
 where  $\nu$ and ${\bm\tau}=\{\tau_1,\cdots,\tau_M \}$ are the corresponding dual variables. Then, there must exist a $\nu^\star$ and ${\bm\tau}^\star=\{\tau_1^\star,\cdots,\tau_M^\star \}$ for ensuring that the following conditions are satisfied:
  \vspace{-0.2cm}
 \begin{eqnarray}
 \!\!\!\!\!\!\!\!\!\!{  {{\nabla _{{\bm{\phi}^*}}}{\cal L}({\bm{\phi}},\nu, \bm{\tau})|_{{\bm{\phi}}={\bm{\phi}}^\star}} }
  \!\!=\!\!{ {{\nabla _{\bm{\phi}^* }}g({\bm{\phi}} |{{\bm{\phi}} ^ \star })|_{\bm{\phi}=\bm{\phi}^ \star} }  }\!\!-\!\!\nu^\star\left( {{\bf{g}^*} \!+\! {\bm{\Upsilon} }{{\bm{\phi}} ^\star}} \right)\!\!+\!\!{\sum\limits_{m = 1}^M \!\!{\tau _m^ \star \!  \left({{\nabla _{\bm{\phi}^* } }\left| {{\phi _m}} \right|}\right)|_{\bm{\phi}=\bm{\phi}^ \star}  }  }\!\!\!&=&\!\!\!{\bf{0}},\label{jyfvtkh}\\
  \!\!\!\!\nu^\star\left( \!\hat Q\!-\!2{\mathop{\rm Re}\nolimits} \left[ \!{{{\bm{\phi}^\star} ^{\rm{H}}}\left( {{\bf{g}^*}\! + \! {\bm{\Upsilon} }{{\bm{\phi}} ^\star}} \right)} \!\right]\!\right) &\!\!\!\!\!=\!\!\!\!\!& 0,\label{fjroe}  \\
   {\tau _m^\star}\left( {\left| {{\phi _m^\star}} \right| - 1} \right) &\!\!\!\!\!=\!\!\!\!\!& 0, \forall m.\label{jfrehoih} \vspace{-0.2cm}
 \end{eqnarray}
According to the second condition of (\ref{adefe}),   we have
 \vspace{-0.2cm}
\begin{equation}\label{wsqdeft}
 {\nabla _{\bm{\phi}^* }}g({\bm{\phi}} |{{\bm{\phi}} ^ \star })|_{{\bm{\phi}}={\bm{\phi}}^\star}=\nabla_{\bm{\phi}^*}f({\bm{\phi}})|_{{\bm{\phi}}={\bm{\phi}}^\star}. \vspace{-0.2cm}
\end{equation}
Upon denoting the OF of Problem  (\ref{appsdsworig}) as  $\varphi({\bm{\phi}})$, which is the same as $f({\bm{\phi}})$ except that $\varphi({\bm{\phi}})$ has more constants, we have
$\nabla_{\bm{\phi}^*}f({\bm{\phi}})|_{{\bm{\phi}}={\bm{\phi}}^\star}=\nabla_{\bm{\phi}^*}\varphi({\bm{\phi}})|_{{\bm{\phi}}={\bm{\phi}}^\star}$.
Combining with (\ref{wsqdeft}), we have ${\nabla _{\bm{\phi}^* }}g({\bm{\phi}} |{{\bm{\phi}} ^ \star })|_{{\bm{\phi}}={\bm{\phi}}^\star}=\nabla_{\bm{\phi}^*}\varphi({\bm{\phi}})|_{{\bm{\phi}}={\bm{\phi}}^\star}$. By substituting it into (\ref{jyfvtkh}), we arrive at
 \vspace{-0.2cm}
\begin{equation}\label{wferge}
\nabla_{\bm{\phi}^*}\varphi({\bm{\phi}})|_{{\bm{\phi}}={\bm{\phi}}^\star}-\nu^\star\left( {{\bf{g}^*} + {\bm{\Upsilon} }{{\bm{\phi}} ^\star}} \right)+{\sum\limits_{m = 1}^M {\tau _m^ \star   \left({{\nabla _{\bm{\phi}^* } }\left| {{\phi _m}} \right|}\right)|_{\bm{\phi}=\bm{\phi}^ \star}  }  }={\bf{0}}. \vspace{-0.2cm}
\end{equation}
It can be checked that the set of equations (\ref{fjroe}), (\ref{jfrehoih}) and (\ref{wferge}) constitutes exactly the  KKT conditions
of Problem  (\ref{appsdsworig}). Hence, the proof is complete.

\section{Proof of Theorem 4}\label{prooftheorem4}

 Let us  define the OF of Problem (\ref{appstneorig}) as
  \vspace{-0.2cm}
\begin{equation}\label{swefrg}
h\left( {{\bf{W}}, {\bf{U}},{\bf{F}},{\bm{\Phi}} } \right)\buildrel \Delta \over= \sum\limits_{k = 1}^{K_I} {{\omega_{k}}{h_{k}}\left( {{\bf{W}}, {\bf{U}},{\bf{F}},{\bm{\Phi}} } \right)}. \vspace{-0.2cm}
\end{equation}
It can be readily verified that the sequence of solutions $\{{{\bf{F}}^{(n)}},{\bm{\phi}}^{(n)}\}$ generated by Algorithm \ref{bcd} is always feasible for Problem (\ref{appstaoneorig}). The monotonic property of Algorithm \ref{bcd} can be similarly proved by using the method of \cite{pancunhua2017}.

In the following, we prove that the converged solution satisfies the KKT conditions of Problem (\ref{appstaoneorig}). Let us denote the converged solution as $\{{\bf{W}}^\star, {\bf{U}}^\star,{\bf{F}}^\star,{\bm{\Phi}}^\star \}$.

According to Theorem 1, ${\bf{F}}^\star$ is the KKT-optimum point of Problem (\ref{appssxsorig}). Upon  denoting the OF of Problem  (\ref{appssxsorig}) as $z({\bf{F}},{\bm\Phi ^ \star })$, the Lagrange function of Problem (\ref{appssxsorig}) is given by
\begin{equation}\label{scsfv}
 {\cal L}({\bf{F}},\lambda,\mu)=z({\bf{F}},{\bm\Phi ^ \star })+\lambda\left(\sum\limits_{k = 1}^{K_I} {\left\| {{{\bf{F}}_{k}}} \right\|_F^2} - {P_{T}}\right)   +\mu \left(\bar Q-    {{\rm{tr}}\left( {\sum\limits_{k = 1}^{{K_I}} {{\bf{F}}_k^{{\rm{H}}}{\bf{G}}{{\bf{F}}_k}} } \right)}  \right),
\end{equation}
where $\lambda$ and $\mu$ are the corresponding dual variables.
Then, there must exist a  $\lambda^\star$ and $\mu^\star$ for ensuring that the following conditions are satisfied \footnote{For simplicity, the prime constraints are omitted.   }:
 \vspace{-0.2cm}
\begin{eqnarray}
  {  {\left. {{\nabla _{{\bf{F}}_k^*}}\cal L({\bf{F}},\lambda ,\mu )} \right|_{{{\bf{F}}_k} = {\bf{F}}_k^ \star }} } \!=\!{  {\left. {{\nabla _{{\bf{F}}_k^*}}z({\bf{F}},{\bm\Phi ^ \star })} \right|_{{{\bf{F}}_k} = {\bf{F}}_k^ \star }} } + \lambda^\star {\bf{F}}_k^\star - \mu^\star {\bf{GF}}_k^\star &\! =\!&  {\bf{0}},\forall k\in {\cal K}_I, \label{jjiofe}  \\
  \lambda^\star\left(\sum\limits_{k = 1}^{K_I} {\left\| {{{\bf{F}}_{k}^\star}} \right\|_F^2} - {P_{T}}\right)  &\!\!=\!\!&0,\label{jjisddsfe}\\
  \mu^\star \left(\bar Q-    {{\rm{tr}}\left( {\sum\limits_{k = 1}^{{K_I}} {{\bf{F}}_k^{\star{\rm{H}}}{\bf{G}}{{\bf{F}}_k^\star}} } \right)}  \right)&\!\!=\!\!& 0. \label{jjissde}
\end{eqnarray}
Furthermore, it can be readily checked that
 \vspace{-0.2cm}
\begin{equation}\label{sefr}
   {\left. {{\nabla _{{{\bf{F}}_k^*}}}h\left( {{{\bf{W}}^ \star },{{\bf{U}}^ \star },{\bf{F}},{\bm\Phi ^ \star }} \right)} \right|_{{{\bf{F}}_k} = {\bf{F}}_k^ \star }} ={\left. {{\nabla _{{{\bf{F}}_k^*}}}z({\bf{F}},{\bm\Phi ^ \star })} \right|_{{{\bf{F}}_k} = {\bf{F}}_k^ \star }}, \forall k \in {\cal K}_I. \vspace{-0.2cm}
\end{equation}
To expound a little further, we have the following chain of inequalities:
 \vspace{-0.2cm}
\begin{align}
 &{\left. {{\nabla _{{{\bf{F}}_k^*}}}h_k\left( {{{\bf{W}}^ \star },{{\bf{U}}^ \star },{\bf{F}},{\bm\Phi ^ \star }} \right)} \right|_{{{\bf{F}}_k} = {\bf{F}}_k^ \star }} \label{hfivehor} \\
 = & - {\rm{tr}}\left( {{\bf{W}}_k^{\star}\left( {{{\left. {{\nabla _{{\bf{F}}_k^*}}{{\bf{E}}_k}\left( {{{\bf{U}}^ \star },{\bf{F}},{\bm\Phi ^ \star }} \right)} \right|}_{{{\bf{F}}_k} = {\bf{F}}_k^ \star }}} \right)} \right)\label{chainrule}\\
  =& - {\rm{tr}}\left( {{{\left( {{{\bf{E}}_k}\left( {{{\bf{U}}^ \star },{{\bf{F}}^ \star },{\bm\Phi ^ \star }} \right)} \right)}^{ - 1}}\left( {{{\left. {{\nabla _{{\bf{F}}_k^*}}{{\bf{E}}_k}\left( {{{\bf{U}}^ \star },{\bf{F}},{\Phi ^ \star }} \right)} \right|}_{{{\bf{F}}_k} = {\bf{F}}_k^ \star }}} \right)} \right)\label{second}\\
 =&{\left. {\left( {{\nabla _{{\bf{F}}_k^*}}\log \left| {{{\left( {{{\bf{E}}_k}\left( {{{\bf{U}}^ \star },{\bf{F}},{\bm\Phi ^ \star }} \right)} \right)}^{ - 1}}} \right|} \right)} \right|_{{{\bf{F}}_k} = {\bf{F}}_k^ \star }}\\
 =& {\left. {{\nabla _{{\bf{F}}_k^*}}{R_k}({\bf{F}},{\bm\Phi ^ \star })} \right|_{{{\bf{F}}_k} = {\bf{F}}_k^ \star }},\label{pvojjnhuy}
\end{align}
where (\ref{chainrule}) follows from the chain rule, and the final equality follows from applying the Woodbury matrix identity to (\ref{xsdvtg}). Combining (\ref{pvojjnhuy}) with (\ref{sefr}), we have
 \vspace{-0.2cm}
\begin{equation}\label{ewfrgt}
 {\left. {{\nabla _{{{\bf{F}}_k^*}}}z({\bf{F}},{\bm\Phi ^ \star })} \right|_{{{\bf{F}}_k} = {\bf{F}}_k^ \star }}={\left. {{\nabla _{{\bf{F}}_k^*}}{R_k}({\bf{F}},{\bm\Phi ^ \star })} \right|_{{{\bf{F}}_k} = {\bf{F}}_k^ \star }}.
\end{equation}
By substituting (\ref{ewfrgt}) into (\ref{jjiofe}), we arrive at
 \vspace{-0.2cm}
 \begin{equation}\label{wdfecvrgth}
{\left. {{\nabla _{{\bf{F}}_k^*}}{R_k}({\bf{F}},{\bm\Phi ^ \star })} \right|_{{{\bf{F}}_k} = {\bf{F}}_k^ \star }} + \lambda^\star {\bf{F}}_k^\star - \mu^\star {\bf{GF}}_k^\star  =  {\bf{0}},\forall k \in {\cal K}_I.
 \end{equation}

According to Theorem 3, ${\bm{\phi}}^\star$ satisfies the KKT conditions of Problem  (\ref{appsdsworig}), and thus the set of equations (\ref{fjroe}), (\ref{jfrehoih}) and (\ref{wferge}) hold.

Furthermore, it can be readily verified that
 \vspace{-0.2cm}
\begin{equation}\label{seedwfr}
   {\nabla _{{\bm{\phi}^*}}}h\left( {\bf{W}}^\star, {\bf{U}}^\star,{\bf{F}}^\star,{\bm{\Phi}} \right) |_{{\bm{\phi}}={\bm{\phi}}^\star} =\nabla_{\bm{\phi}^*}\varphi({\bm{\phi}})|_{{\bm{\phi}}={\bm{\phi}}^\star}.
\end{equation}
By using  similar derivations as in (\ref{hfivehor})-(\ref{pvojjnhuy}), we can prove that
 \vspace{-0.2cm}
\begin{equation}\label{wdef}
 {\nabla _{{\bm{\phi}^*}}}h\left( {\bf{W}}^\star, {\bf{U}}^\star,{\bf{F}}^\star,{\bm{\Phi}} \right) |_{{\bm{\phi}}={\bm{\phi}}^\star}= {\left. {{\nabla _{\bm{\phi}^*}}{R_k}(\bm{\phi},{\bf{F}}^\star)} \right|_{{\bm{\phi}} =\bm{\phi}^ \star }}.
\end{equation}
Hence, we have
\begin{equation}\label{trtuki}
  \nabla_{\bm{\phi}^*}\varphi({\bm{\phi}})|_{{\bm{\phi}}={\bm{\phi}}^\star}={\left. {{\nabla _{\bm{\phi}^*}}{R_k}(\bm{\phi},{\bf{F}}^\star)} \right|_{{\bm{\phi}} =\bm{\phi}^ \star }}.
\end{equation}
By substituting (\ref{trtuki}) into (\ref{wferge}), we arrive at:
\begin{equation}\label{stgb}
{\left. {{\nabla _{\bm{\phi}^*}}{R_k}(\bm{\phi},{\bf{F}}^\star)} \right|_{{\bm{\phi}} =\bm{\phi}^ \star }}-\nu^\star\left( {{\bf{g}^*} \!+\! {\bm{\Upsilon} }{{\bm{\phi}} ^\star}} \right)+{\sum\limits_{m = 1}^M {\tau _m^ \star \!  \left({{\nabla _{\bm{\phi}^* } }\left| {{\phi _m}} \right|}\right)|_{\bm{\phi}=\bm{\phi}^ \star}  }  }={\bf{0}}.
\end{equation}

Then, the set of equations (\ref{wdfecvrgth}), (\ref{jjisddsfe}), (\ref{jjissde}), (\ref{stgb}), (\ref{fjroe}), and (\ref{jfrehoih}) constitute exactly the KKT conditions of Problem (\ref{appstaoneorig}).

\end{appendices}

\
\



%\bibliographystyle{IEEEtran}
%% argument is your BibTeX string definitions and bibliography database(s)
%\bibliography{myre}


\vspace{-0.5cm}
%\vspace{-1.2cm}
\bibliographystyle{IEEEtran}
\bibliography{myre}

\begin{thebibliography}{10}
\providecommand{\url}[1]{#1}
\csname url@samestyle\endcsname
\providecommand{\newblock}{\relax}
\providecommand{\bibinfo}[2]{#2}
\providecommand{\BIBentrySTDinterwordspacing}{\spaceskip=0pt\relax}
\providecommand{\BIBentryALTinterwordstretchfactor}{4}
\providecommand{\BIBentryALTinterwordspacing}{\spaceskip=\fontdimen2\font plus
\BIBentryALTinterwordstretchfactor\fontdimen3\font minus
  \fontdimen4\font\relax}
\providecommand{\BIBforeignlanguage}[2]{{%
\expandafter\ifx\csname l@#1\endcsname\relax
\typeout{** WARNING: IEEEtran.bst: No hyphenation pattern has been}%
\typeout{** loaded for the language `#1'. Using the pattern for}%
\typeout{** the default language instead.}%
\else
\language=\csname l@#1\endcsname
\fi
#2}}
\providecommand{\BIBdecl}{\relax}
\BIBdecl

\bibitem{di2019smart}
M.~Di~Renzo, M.~Debbah, D.-T. Phan-Huy, A.~Zappone, M.-S. Alouini, C.~Yuen,
  V.~Sciancalepore, G.~C. Alexandropoulos, J.~Hoydis, H.~Gacanin \emph{et~al.},
  ``Smart radio environments empowered by reconfigurable {AI} meta-surfaces: an
  idea whose time has come,'' \emph{EURASIP J. Wireless Commun. Networking},
  vol. 2019, no.~1, p. 129, 2019.

\bibitem{qingqing2019towards}
\BIBentryALTinterwordspacing
Q.~Wu and R.~Zhang, ``Towards smart and reconfigurable environment: Intelligent
  reflecting surface aided wireless network.'' [Online]. Available:
  \url{https://arxiv.org/abs/1905.00152}
\BIBentrySTDinterwordspacing

\bibitem{zhang2019multiple}
\BIBentryALTinterwordspacing
J.~Zhang, E.~Bj{\"o}rnson, M.~Matthaiou, D.~W.~K. Ng, H.~Yang, and D.~Love,
  ``Multiple antenna technologies for beyond {5G}.'' [Online]. Available:
  \url{https://arxiv.org/abs/1910.00092}
\BIBentrySTDinterwordspacing

\bibitem{cui2014coding}
T.~J. Cui, M.~Q. Qi, X.~Wan, J.~Zhao, and Q.~Cheng, ``Coding metamaterials,
  digital metamaterials and programmable metamaterials,'' \emph{Light: Science
  \& Applications}, vol.~3, no.~10, p. e218, 2014.

\bibitem{zhang2019cell}
J.~Zhang, S.~Chen, Y.~Lin, J.~Zheng, B.~Ai, and L.~Hanzo, ``Cell-free massive
  {MIMO}: A new next-generation paradigm,'' \emph{IEEE Access}, vol.~7, pp.
  99\,878--99\,888, 2019.

\bibitem{zhang2018mixed}
J.~Zhang, L.~Dai, Z.~He, B.~Ai, and O.~A. Dobre, ``Mixed-{ADC/DAC} multipair
  massive {MIMO} relaying systems: Performance analysis and power
  optimization,'' \emph{IEEE Trans. Commun.}, vol.~67, no.~1, pp. 140--153,
  2018.

\bibitem{qingqingwuglobe}
Q.~{Wu} and R.~{Zhang}, ``Intelligent reflecting surface enhanced wireless
  network: Joint active and passive beamforming design,'' in \emph{2018 IEEE
  Global Communications Conference (GLOBECOM)}, Dec. 2018, pp. 1--6.

\bibitem{yu2019miso}
\BIBentryALTinterwordspacing
X.~Yu, D.~Xu, and R.~Schober, ``{MISO} wireless communication systems via
  intelligent reflecting surfaces.'' [Online]. Available:
  \url{https://arxiv.org/abs/1904.12199}
\BIBentrySTDinterwordspacing

\bibitem{yang2019intelligent}
\BIBentryALTinterwordspacing
Y.~Yang, B.~Zheng, S.~Zhang, and R.~Zhang, ``Intelligent reflecting surface
  meets {OFDM}: Protocol design and rate maximization.'' [Online]. Available:
  \url{https://arxiv.org/abs/1906.09956}
\BIBentrySTDinterwordspacing

\bibitem{han2019large}
Y.~Han, W.~Tang, S.~Jin, C.~Wen, and X.~Ma, ``Large intelligent
  surface-assisted wireless communication exploiting statistical {CSI},''
  \emph{IEEE Trans. Veh. Technol.}, 2019.

\bibitem{abeywickrama2019intelligent}
\BIBentryALTinterwordspacing
S.~Abeywickrama, R.~Zhang, and C.~Yuen, ``Intelligent reflecting surface:
  Practical phase shift model and beamforming optimization.'' [Online].
  Available: \url{https://arxiv.org/abs/1907.06002}
\BIBentrySTDinterwordspacing

\bibitem{wu2018intelligent}
Q.~{Wu} and R.~{Zhang}, ``Intelligent reflecting surface enhanced wireless
  network via joint active and passive beamforming,'' \emph{IEEE Trans.
  Wireless Commun.}, vol.~18, no.~11, pp. 5394--5409, Nov. 2019.

\bibitem{huang2019reconfigurable}
C.~{Huang}, A.~{Zappone}, G.~C. {Alexandropoulos}, M.~{Debbah}, and C.~{Yuen},
  ``Reconfigurable intelligent surfaces for energy efficiency in wireless
  communication,'' \emph{IEEE Trans. Wireless Commun.}, vol.~18, no.~8, pp.
  4157--4170, Aug. 2019.

\bibitem{guo2019weighted}
\BIBentryALTinterwordspacing
H.~Guo, Y.~Liang, J.~Chen, and E.~Larsson, ``Weighted sum-rate optimization for
  intelligent reflecting surface enhanced wireless networks.'' [Online].
  Available: \url{https://arxiv.org/abs/1905.07920}
\BIBentrySTDinterwordspacing

\bibitem{nadeem2019large}
\BIBentryALTinterwordspacing
Q.~Nadeem, A.~Kammoun, A.~Chaaban, M.~Debbah, and M.~Alouini, ``Large
  intelligent surface assisted {MIMO} communications.'' [Online]. Available:
  \url{https://arxiv.org/abs/1903.08127}
\BIBentrySTDinterwordspacing

\bibitem{yu2019enabling}
\BIBentryALTinterwordspacing
X.~Yu, D.~Xu, and R.~Schober, ``Enabling secure wireless communications via
  intelligent reflecting surfaces.'' [Online]. Available:
  \url{https://arxiv.org/abs/1904.09573}
\BIBentrySTDinterwordspacing

\bibitem{cui2019secure}
M.~Cui, G.~Zhang, and R.~Zhang, ``Secure wireless communication via intelligent
  reflecting surface,'' \emph{IEEE Wireless Commun. Lett.}, 2019.

\bibitem{hongshen}
H.~{Shen}, W.~{Xu}, W.~{Xu}, S.~{Gong}, Z.~{He}, and C.~{Zhao}, ``Secrecy rate
  maximization for intelligent reflecting surface assisted multi-antenna
  communications,'' \emph{IEEE Commun. Lett.}, pp. 1--1, 2019.

\bibitem{chen2019intelligent}
\BIBentryALTinterwordspacing
J.~Chen, Y.~Liang, Y.~Pei, and H.~Guo, ``Intelligent reflecting surface: A
  programmable wireless environment for physical layer security.'' [Online].
  Available: \url{https://arxiv.org/abs/1905.03689}
\BIBentrySTDinterwordspacing

\bibitem{xu2019resource}
\BIBentryALTinterwordspacing
D.~Xu, X.~Yu, Y.~Sun, D.~W.~K. Ng, and R.~Schober, ``Resource allocation for
  secure {IRS}-assisted multiuser {MISO} systems.'' [Online]. Available:
  \url{https://arxiv.org/abs/1907.03085}
\BIBentrySTDinterwordspacing

\bibitem{guan2019intelligent}
\BIBentryALTinterwordspacing
X.~Guan, Q.~Wu, and R.~Zhang, ``Intelligent reflecting surface assisted secrecy
  communication via joint beamforming and jamming.'' [Online]. Available:
  \url{https://arxiv.org/abs/1907.12839}
\BIBentrySTDinterwordspacing

\bibitem{tong2019}
\BIBentryALTinterwordspacing
T.~Bai, C.~Pan, Y.~Deng, M.~Elkashlan, A.~Nallanathan, and L.~Hanzo, ``Latency
  minimization for intelligent reflecting surface aided mobile edge
  computing.'' [Online]. Available: \url{https://arxiv.org/abs/1910.07990}
\BIBentrySTDinterwordspacing

\bibitem{gui2019}
\BIBentryALTinterwordspacing
G.~Zhou, C.~Pan, H.~Ren, K.~Wang, and A.~Nallanathan, ``Intelligent reflecting
  surface aided multigroup multicast {MISO} communication systems.'' [Online].
  Available: \url{https://arxiv.org/abs/1909.04606}
\BIBentrySTDinterwordspacing

\bibitem{pan2019intelligent}
\BIBentryALTinterwordspacing
C.~Pan, H.~Ren, K.~Wang, W.~Xu, M.~Elkashlan, A.~Nallanathan, and L.~Hanzo,
  ``Intelligent reflecting surface for multicell {MIMO} communications.''
  [Online]. Available: \url{https://arxiv.org/abs/1907.10864}
\BIBentrySTDinterwordspacing

\bibitem{huang2019indoor}
\BIBentryALTinterwordspacing
C.~Huang, G.~C. {Alexandropoulos}, C.~Yuan, and M.~{Debbah}, ``Indoor signal
  focusing with deep learning designed reconfigurable intelligent surfaces.''
  [Online]. Available: \url{https://arxiv.org/abs/1905.07726}
\BIBentrySTDinterwordspacing

\bibitem{he2019cascaded}
\BIBentryALTinterwordspacing
Z.~He and X.~Yuan, ``Cascaded channel estimation for large intelligent
  metasurface assisted massive {MIMO}.'' [Online]. Available:
  \url{https://arxiv.org/abs/1905.07948}
\BIBentrySTDinterwordspacing

\bibitem{taha2019enabling}
\BIBentryALTinterwordspacing
A.~Taha, M.~Alrabeiah, and A.~Alkhateeb, ``Enabling large intelligent surfaces
  with compressive sensing and deep learning.'' [Online]. Available:
  \url{https://arxiv.org/abs/1904.10136}
\BIBentrySTDinterwordspacing

\bibitem{zhou2019robust}
\BIBentryALTinterwordspacing
G.~Zhou, C.~Pan, H.~Ren, K.~Wang, M.~D. Renzo, and A.~Nallanathan, ``Robust
  beamforming design for intelligent reflecting surface aided {MISO}
  communication systems.'' [Online]. Available:
  \url{https://arxiv.org/abs/1911.06237}
\BIBentrySTDinterwordspacing

\bibitem{zhang2013mimo}
R.~Zhang and C.~K. Ho, ``{MIMO} broadcasting for simultaneous wireless
  information and power transfer,'' \emph{IEEE Trans. Wireless Commun.},
  vol.~12, no.~5, pp. 1989--2001, May 2013.

\bibitem{wu2019weighted}
\BIBentryALTinterwordspacing
Q.~Wu and R.~Zhang, ``Weighted sum power maximization for intelligent
  reflecting surface aided {SWIPT}.'' [Online]. Available:
  \url{https://arxiv.org/abs/1907.05558}
\BIBentrySTDinterwordspacing

\bibitem{yansun}
Y.~{Sun}, P.~{Babu}, and D.~P. {Palomar}, ``Majorization-minimization
  algorithms in signal processing, communications, and machine learning,''
  \emph{IEEE Trans. Signal Process.}, vol.~65, no.~3, pp. 794--816, Feb. 2017.

\bibitem{cunhua2016}
C.~{Pan}, W.~{Xu}, J.~{Wang}, H.~{Ren}, W.~{Zhang}, N.~{Huang}, and M.~{Chen},
  ``Pricing-based distributed energy-efficient beamforming for {MISO}
  interference channels,'' \emph{IEEE J. Sel. Areas Commun.}, vol.~34, no.~4,
  pp. 710--722, Apr. 2016.

\bibitem{pancunhua2017}
C.~{Pan}, H.~{Zhu}, N.~J. {Gomes}, and J.~{Wang}, ``Joint precoding and {RRH}
  selection for user-centric green {MIMO C-RAN},'' \emph{IEEE Trans. Wireless
  Commun.}, vol.~16, no.~5, pp. 2891--2906, May 2017.

\bibitem{boshkovska2015practical}
E.~Boshkovska, D.~W.~K. Ng, N.~Zlatanov, and R.~Schober, ``Practical non-linear
  energy harvesting model and resource allocation for {SWIPT} systems,''
  \emph{IEEE Commun. Lett.}, vol.~19, no.~12, pp. 2082--2085, 2015.

\bibitem{mishra2018energy}
D.~Mishra, G.~C. Alexandropoulos, and S.~De, ``Energy sustainable {IoT} with
  individual {QoS} constraints through {MISO} {SWIPT} multicasting,''
  \emph{IEEE Internet Things J.}, vol.~5, no.~4, pp. 2856--2867, Aug. 2018.

\bibitem{xiong2017rate}
K.~Xiong, B.~Wang, and K.~R. Liu, ``Rate-energy region of {SWIPT} for {MIMO}
  broadcasting under nonlinear energy harvesting model,'' \emph{IEEE Trans.
  Wireless Commun.}, vol.~16, no.~8, pp. 5147--5161, Aug. 2017.

\bibitem{fangwang2018}
F.~{Wang}, J.~{Xu}, X.~{Wang}, and S.~{Cui}, ``Joint offloading and computing
  optimization in wireless powered mobile-edge computing systems,'' \emph{IEEE
  Trans. Wireless Commun.}, vol.~17, no.~3, pp. 1784--1797, Mar. 2018.

\bibitem{xu2014multiuser}
J.~Xu, L.~Liu, and R.~Zhang, ``Multiuser {MISO} beamforming for simultaneous
  wireless information and power transfer,'' \emph{IEEE Trans. Signal
  Process.}, vol.~62, no.~18, pp. 4798--4810, Sep. 2014.

\bibitem{shi2011iteratively}
Q.~Shi, M.~Razaviyayn, Z.-Q. Luo, and C.~He, ``An iteratively weighted {MMSE}
  approach to distributed sum-utility maximization for a {MIMO} interfering
  broadcast channel,'' \emph{IEEE Trans. Signal Process.}, vol.~59, no.~9, pp.
  4331--4340, Sep. 2011.

\bibitem{cunhuajsac}
C.~{Pan}, H.~{Ren}, M.~{Elkashlan}, A.~{Nallanathan}, and L.~{Hanzo}, ``The
  non-coherent ultra-dense {C-RAN} is capable of outperforming its coherent
  counterpart at a limited fronthaul capacity,'' \emph{IEEE J. Sel. Areas
  Commun.}, vol.~36, no.~11, pp. 2549--2560, Nov. 2018.

\bibitem{pan2018robust}
\BIBentryALTinterwordspacing
C.~Pan, H.~Ren, M.~Elkashlan, A.~Nallanathan, and L.~Hanzo, ``Robust
  beamforming design for ultra-dense user-centric {C-RAN} in the face of
  realistic pilot contamination and limited feedback.'' [Online]. Available:
  \url{https://arxiv.org/abs/1804.03990}
\BIBentrySTDinterwordspacing

\bibitem{grant2014cvx}
M.~Grant and S.~Boyd, ``{CVX}: Matlab software for disciplined convex
  programming, version 2.1,'' 2014.

\bibitem{boyd2004convex}
S.~Boyd and L.~Vandenberghe, \emph{Convex optimization}.\hskip 1em plus 0.5em
  minus 0.4em\relax Cambridge university press, 2004.

\bibitem{cunhuawcl}
C.~{Pan}, W.~{Xu}, W.~{Zhang}, J.~{Wang}, H.~{Ren}, and M.~{Chen}, ``Weighted
  sum energy efficiency maximization in ad hoc networks,'' \emph{IEEE Wireless
  Commun. Lett.}, vol.~4, no.~3, pp. 233--236, Jun. 2015.

\bibitem{zhang2017matrix}
X.-D. Zhang, \emph{Matrix analysis and applications}.\hskip 1em plus 0.5em
  minus 0.4em\relax Cambridge University Press, 2017.

\bibitem{jiansong}
J.~{Song}, P.~{Babu}, and D.~P. {Palomar}, ``Sequence design to minimize the
  weighted integrated and peak sidelobe levels,'' \emph{IEEE Trans. Signal
  Process.}, vol.~64, no.~8, pp. 2051--2064, Apr. 2016.

\end{thebibliography}


\end{document}